\documentclass[aps,prb,amsmath,amsfonts,amssymb,showpacs,superscriptaddress,floatfix,twocolumn]{revtex4-1}
\usepackage{times}
\usepackage{epsfig}
\usepackage{hyperref}
\usepackage{bm}
\usepackage{color}
\newcommand{\mods}{\color[rgb]{1,0,0}} % Modif. start
\newcommand{\mode}{\color[rgb]{0,0,0}} % Modif. end
\begin{document}
%%%%%%%%%%%%%%%%%%%%%%%%%%%%%%%%%%%%%%%%%%%%%%%%%%%%%%%%%%%%%%%%%%%%%%%%%%%%%%%%%
\title{Dynamic Peierls-Nabarro equations for elastically isotropic crystals}
%~~~~~~~~~~~~~~~~~~~~~~~~~~~~~~~~~~~~~~~~~~~~~~~~~~~~~~~~~~~~~~~~~~~~~~~~~~~~~~~~
\author{Yves-Patrick \surname{Pellegrini}}
\email{yves-patrick.pellegrini@cea.fr}
\affiliation{CEA, DAM, DIF, F-91297 Arpajon, France}
%~~~~~~~~~~~~~~~~~~~~~~~~~~~~~~~~~~~~~~~~~~~~~~~~~~~~~~~~~~~~~~~~~~~~~~~~~~~~~~~~
\date{Received 13 August 2009; revised manuscript received 25 November 2009; published 5 January 2010}
%~~~~~~~~~~~~~~~~~~~~~~~~~~~~~~~~~~~~~~~~~~~~~~~~~~~~~~~~~~~~~~~~~~~~~~~~~~~~~~~~
\begin{abstract}
The dynamic generalization of the Peierls-Nabarro equation for dislocations cores in an isotropic elastic medium is derived for screw, and edge dislocations of the ``glide" and ``climb" type, by means of Mura's eigenstrains method. These equations are of the integro-differential type and feature a non-local kernel in space and time. The equation for the screw differs by an instantaneous term from a previous attempt by Eshelby. Those for both types of edges involve in addition an unusual convolution with the \emph{second} spatial derivative of the displacement jump. As a check, it is shown that these equations correctly reduce, in the stationary limit and for all three types of dislocations, to Weertman's equations that extend the static Peierls-Nabarro model to finite constant velocities.
\end{abstract}
%~~~~~~~~~~~~~~~~~~~~~~~~~~~~~~~~~~~~~~~~~~~~~~~~~~~~~~~~~~~~~~~~~~~~~~~~~~~~~~~~
ArXiv version of: {\small PHYSICAL REVIEW B} \textbf{81}, 024101\hspace{-0.49ex} (2010). \hfill\mods{Minor typos in PRB edition corrected in red.}\mode
%~~~~~~~~~~~~~~~~~~~~~~~~~~~~~~~~~~~~~~~~~~~~~~~~~~~~~~~~~~~~~~~~~~~~~~~~~~~~~~~~
\maketitle
%%%%%%%%%%%%%%%%%%%%%%%%%%%%%%%%%%%%%%%%%%%%%%%%%%%%%%%%%%%%%%%%%%%%%%%%%%%%%%%%
%%%%%%%%%%%%%%%%%%%%%%%%%%%%%%%%%%%%%%%%%%%%%%%%%%%%%%%%%%%%%%%%%%%%%%%%%%%%%%%%
\section{Introduction}
%%%%%%%%%%%%%%%%%%%%%%%%%%%%%%%%%%%%%%%%%%%%%%%%%%%%%%%%%%%%%%%%%%%%%%%%%%%%%%%%
Plastic deformation in crystals occurs as dislocations move through the material under an applied stress.\cite{HIRT82,MURA87} Major quantitative progresses in plasticity modeling arose with the outbreak of the Peierls-Nabarro (PN) integral equation.\cite{PEIE40,NABA47,ESHE49} Aimed at computing dislocation core shapes, this equation establishes a quantitative link between atomic forces, nowadays described by means of the material-dependent $\gamma$ potential (a ``reduced" lattice potential specialized to shear deformations)\cite{CHRI70} and the dislocation core structure. Since, numerous refinements of various nature improved the agreement between the PN model and molecular statics simulations, though best matches with experiment for the core width and the Peierls stress\cite{PEIE40} are obtained so far not by using the PN model, but by addressing \textit{ab initio} the full three-dimensional structure of dislocations cores. Still, in spite of known drawbacks, the PN equation remains widely used. These questions have recently been reviewed by Schoeck.\cite{SCHO05}

Yet, the \emph{dynamic} instance of the Peierls-Nabarro equation appears as a long-standing elusive issue in dislocation theory. To date, simulations (using molecular dynamics\cite{MARI04,VAND04,OLMS05} or phase-field methods\cite{WANG01}) constitute the privileged path to specific dynamic core-related phenomena. Among the latter are the long-hypothesized transonic or supersonic transitions,\cite{FRAN49,ESHE53,ESHE56,WEER67,WEER69} first reported in atomistic simulations\cite{GUMB99,LI02,MARI06,TSUZ08} but only recently exhibited experimentally (in a two-dimensional plasma crystal).\cite{NOSE07} However, their current understanding is far from complete, given the variety of situations that can arise depending on the nature of the crystal and environmental conditions. Analytical progresses have also been made in this direction, but most often at the price of restrictive approximations for the dislocation core.\cite{MARK80,CALL80,SHAR06,MARK08} It should be noted that the major difficulty involved in sonic transitions in real crystals, recognized long ago,\cite{ESHE56} resides in that a dislocation can be part subsonic and part supersonic at the same time due to spatial-dispersion effects.\cite{MARI06} Tackling this problem is not attempted here. However, in the above context and even within a non-dispersive approximation, a one-dimensional dynamic PN equation that leaves all freedom to the core shape would certainly constitute a useful additional tool.

The truth is, in his classical '53 paper on the dynamic motion of dislocations,\cite{ESHE53} Eshelby did write down a dynamic generalization of the PN equation for the screw dislocation. However, acknowledging its complexity he did not use it, focusing instead on an equation of motion for screw dislocations under an assumption of rigid core. A dynamic PN equation for edge dislocations was never proposed, and in practice the only velocity-dependent PN equations studied so far are Weertman's\cite{WEER69} and its modifications,\cite{ROSA01} which apply to \emph{constant} velocities only. Despite a number of analytical explorations of the dynamic regime, this gap has not been filled in yet.

Quite unexpectedly, a close examination of Eshelby's dynamic equation for screws\cite{ESHE53} leads one to conclude that \emph{it does not reduce to Weertman's equation in the stationary limit}. This can be seen from the calculations in Appendix \mods C\mode1. One clue to the reason of this discrepancy is provided by the recent observation\cite{PELL09} that classical static expressions for dislocation-generated displacements, such as that found in Refs.~\onlinecite{HIRT82,MURA87}, miss one term (a distribution) that represents the non-elastically relaxed slip. This term proves irrelevant to the standard static PN model, and cannot be spotted from the static elastic strains alone, since it preserves registry (Sec.\ \ref{sec:proform}). It is shown below that one term of similar origin \emph{is} relevant to dynamic calculations, and provides the explanation for the above discrepancy. With this observation, the Green's function machinery\cite{MURA87} can safely be harnessed to produce the desired dynamic PN equations for screws and edges, that correctly admit Weertman's equations as stationary limits, provided that attention is paid to distributional parts in carrying out various Fourier integrals (Sec.\ \ref{sec:dynPN}).

For convenience, indices $i=x$, $y$, and $z$ or $1$, $2$, and $3$ are used indifferently hereafter. To ease calculations, most of the integrals are read in Ref.~\onlinecite{GRAD07}. Reference is made to these integrals by their book classification number, preceded by ``G.R".

%%%%%%%%%%%%%%%%%%%%%%%%%%%%%%%%%%%%%%%%%%%%%%%%%%%%%%%%%%%%%%%%%%%%%%%%%%%%%%%%
\section{\label{sec:proform}Green's function approach to dislocations}
%%%%%%%%%%%%%%%%%%%%%%%%%%%%%%%%%%%%%%%%%%%%%%%%%%%%%%%%%%%%%%%%%%%%%%%%%%%%%%%%
\subsection{Eigenstrains and dynamic Green's function}
%-------------------------------------------------------------------------------
Inclusions or defects such as dislocations produce distortions in their surrounding medium. The total distortion $\bm{\beta}$ is the gradient of the material displacement $\mathbf{u}$, such that $\beta_{ij}(\mathbf{x})=\partial_j u_i$. Its symmetric part is the total strain $\varepsilon_{ij}=(1/2)(\beta_{ij}+\beta_{ji})$. Assuming small deformations, the total distortion produced by a defect can be written as the sum of a linear elastic distortion $\bm{\beta}^e$, and of a ``nonlinear" part $\bm{\beta}^*$ usually called \emph{eigendistortion}, \cite{MURA87,KRON58}
$\beta_{ij}=\beta^e_{ij}+\beta^*_{ij}$, none of the latter quantities being a gradient in general. Whereas the eigendistortion represents a purely geometric, rigid, i.e., non-elastically relaxed,\cite{NOTE0} contribution to the total distortion that results from the insertion of the inclusion, the elastic distortion represents the elastic relaxation correction that confers to $\bm{\beta}$ a gradient character. A similar decomposition holds for the strain:
$\varepsilon_{ij}=\varepsilon^e_{ij}+\varepsilon^*_{ij}$.

The Green's function approach to dislocations
\cite{MURA87} consists in representing the dislocation by an eigendistortion (localized on a the glide plane) whose physical interpretation is given below (Sec.\ \ref{sec:dperm}) and in computing the induced displacement field $\mathbf{u}$ using an elementary solution of the equations of elasticity. The total distortion  $\beta_{ij}$ ensues, from which $\bm{\beta}^e$ is obtained by subtracting  $\bm{\beta}^*$. Finally, the stress
$\bm{\sigma}$ follows from the linear-elastic strain $\bm{\varepsilon}^e=\mathop{\text{sym}}\bm{\beta}^e$ and linear elasticity, as
\begin{equation}
\label{eq:stress}
\sigma_{ij}=C_{ijkl}\varepsilon^e_{kl}=C_{ijkl}(\varepsilon_{kl}-\varepsilon^*_{kl})=C_{ijkl}(\partial_k
u_l-\beta^*_{kl}),
\end{equation}
where $C_{ijkl}=C_{ijlk}=C_{klij}$ are components of the elastic tensor. Momentum conservation in the form
$\partial_j\sigma_{ij}=\rho\partial_t^2 u_i$, where $\rho$ is the mass density, is written as
%\cite{BROS64}
\begin{equation}
\label{eq:eqgen} C_{ijkl}\partial_j\partial_k u_l-\rho\partial_t^2
u_i=\partial_j \tau_{ij},
\end{equation}
where $\tau_{ij}\equiv C_{ijkl}\beta^*_{kl}$. In an infinite medium, the Green's function of the displacement, $\mathsf{G}(\mathbf{x},t)$, is the solution corresponding to a point-like source located at the origin of space and time (the minus sign is conventional),\cite{MURA87}
\begin{equation}
\label{eq:momentum}
C_{ijkl}\partial_j\partial_k
G_{lm}(\mathbf{x},t)-\rho\partial_t^2G_{im}(\mathbf{x},t)=-\delta_{im}\delta(\mathbf{x})\delta(t).
\end{equation}

The following space-time Fourier transform (FT) convention is used hereafter ($f$ is an arbitrary function):
\begin{equation}
f(\mathbf{x},t)=\int
\frac{\mathrm{d}^3\!k}{(2\pi)^3}\frac{\mathrm{d}\omega}{2\pi}f(\mathbf{k},\omega)e^{i(\mathbf{k}\cdot\mathbf{x}-\omega
t )}.
\end{equation}
Introducing the acoustic tensor $\mathsf{N}$
of components $N_{ij}=C_{iklj}k_kk_l$ and the identity matrix $\mathsf{I}$ of components $\delta_{ij}$, the solution to Eq.\ (\ref{eq:momentum}) reads in matrix notation
\begin{equation}
\label{eq:ninv}
\mathsf{G}(\mathbf{k},\omega)=\left(\mathsf{N}-\rho\omega^2\mathsf{I}\right)^{-1}.
\end{equation}
It is convenient for the problem at hand to work in the mixed ``space Fourier modes/time" representation. By convolution of the elementary solution, the solution to Eq.\ (\ref{eq:eqgen}) is obtained as
\begin{equation}
\label{eq:udyn}
u_i(\mathbf{x},t)=-i
\int\hspace{-1ex}\mathrm{d}t'\hspace{-1ex}
\int\hspace{-1ex}\frac{\mathrm{d}^3k}{(2\pi)^3}
G_{ij}(\mathbf{k},t-t')k_k\tau_{jk}(\mathbf{k},t')\,e^{i\mathbf{k}\cdot\mathbf{x}},
\end{equation}
where the integrals run from $-\infty$ to $+\infty$,
and the stress follows from Eq.\ (\ref{eq:stress}). Henceforth, we confine ourselves to the simplest isotropic case for which
\begin{equation}
\label{eq:elast}
C_{ijkl}=\lambda\,\delta_{ij}\delta_{kl}+\mu(\delta_{ik}\delta_{jl}+\delta_{il}\delta_{jk}),
\end{equation}
where $\mu$ is the shear modulus and $\lambda$ is the Lam\'e coefficient. Introduce moreover the shear and longitudinal sound velocities
$c_{\text{S}}=\sqrt{\mu/\rho}$ and $c_{\text{L}}=\sqrt{(\lambda+2\mu)/\rho}$. The inverse in Eq.\ (\ref{eq:ninv}) is immediate in the basis of longitudinal and transverse projectors with respect to $\mathbf{\hat k}=\mathbf{k}/k$. Thus,
\begin{equation}
\mathsf{N}=\mu k^2\left[(\mathsf{I}-\mathbf{\hat k}\mathbf{\hat
k})+(c_{\text{L}}/c_{\text{S}})^2\mathbf{\hat k}\mathbf{\hat k}\right]
\end{equation}
and the dynamic Green's function of the displacement reads
\begin{eqnarray}
\mathsf{G}(\mathbf{k},\omega)&=&
\left(\mathsf{N}-\rho\omega^2\mathsf{I}\right)^{-1}\nonumber\\
\label{eq:gdym}
&=&\frac{1}{\mu}
\left[\frac{\mathsf{I}-\mathbf{\hat k}\mathbf{\hat
k}}{k^2-(\omega/c_{\text{S}})^2}+\frac{c_{\text{S}}^2}{c_{\text{L}}^2}\frac{\mathbf{\hat
k}\mathbf{\hat k}}{k^2-(\omega/c_{\text{L}})^2}\right].
\end{eqnarray}
Its static limit is more conveniently expressed in terms of the Poisson ratio $\nu=\lambda/[2(\lambda+\mu)]$, such that
$c_{\text{S}}^2/c_{\text{L}}^2=(1-2\nu)/[2(1-\nu)]$,
\begin{equation}
\label{eq:gstat} G_{ij}(\mathbf{k})\equiv
G_{ij}(\mathbf{k},\omega=0)=\frac{1}{\mu
k^2}\left[\delta_{ij}-\frac{\hat{k}_i\hat{k}_j}{2(1-\nu)}\right].
\end{equation}
Let $\theta(x)$ denote the Heaviside function. Inverting the time-Fourier transform in Eq.\ (\ref{eq:gdym}) with a suitable choice of contour in the $\omega$-complex plane\cite{MORS53} yields the following retarded Green's function, which describes waves going away from the source:
\begin{eqnarray}
G_{ij}(\mathbf{k},t)&=&\frac{\theta(t)c_{\text{S}}^2}{\mu
k}\Bigl[\frac{1}{c_{\text{S}}}\sin(c_{\text{S}} k
t)(\delta_{ij}-\hat{k}_i\hat{k}_j)\nonumber\\
\label{eq:gdyn}
&&\hspace{1.5cm}{}+\frac{1}{c_{\text{L}}}\sin(c_{\text{L}} k t)\hat{k}_i\hat{k}_j\Bigr].
\end{eqnarray}

%-------------------------------------------------------------------------------
\subsection{Volterra dislocations and importance of history}
\label{sec:dperm}
%-------------------------------------------------------------------------------
The problem of finding the fields associated to a dislocation with an extended core is most efficiently split up in two steps. First, a solution is derived for a Volterra dislocation of infinitely narrow core, which only slightly complicates the above calculation for a pointlike source. In the second step, a convolution product of the obtained elementary solution with the shape of the extended core, considered as a superposition of Volterra dislocations, is taken according to the superposition principle of solutions of linear elasticity. This approach, introduced by Eshelby,\cite{ESHE49} is well suited to obtaining the Peierls-Nabarro integral equation. Indeed in this equation (one of stress balance) where the core shape itself is the unknown, the convolution integral cannot be explicitly evaluated in general.

\begin{figure}
\includegraphics[width=8cm]{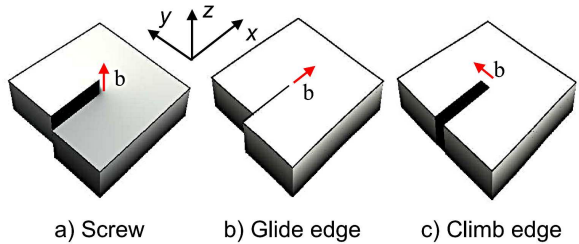}
\caption{\label{fig:fig1}(color online) Three different types of Volterra dislocations (core of null width), as computed using Eqs.\ (\ref{eq:u3ok}), (\ref{eq:ufieldge-stat}) and (\ref{eq:ufieldgec-stat}) with $\nu=0.3$, $\varepsilon_g=10^{-4}$ and $b=1$ ($|x|$, $|y|\leq 5$) (Ref.\ \onlinecite{NOTE1}}).
\end{figure}

The eigendistortions associated to the three relevant types of rectilinear infinite Volterra dislocations, with dislocation line along the $Oz$ axis, are represented as follows (the core lies at the origin of the Cartesian axes):
\begin{subequations}
\label{eq:betastars}
\begin{eqnarray}
\label{eq:vis}
\beta^*_{ij}(\mathbf{x})&=&b\,\delta(y)\theta(-x)\delta_{i3}\delta_{j2}\qquad
(\text{screw}),\\
\label{eq:coing}
\beta^*_{ij}(\mathbf{x})&=&b\,\delta(y)\theta(-x)\delta_{i1}\delta_{j2}\qquad
(\text{glide edge}),\\
\label{eq:coins}
\beta^*_{ij}(\mathbf{x})&=&b\,\delta(y)\theta(-x)\delta_{i2}\delta_{j2}\qquad
(\text{climb edge}).
\end{eqnarray}
\end{subequations}
The norm of the Burgers vector $\mathbf{b}$ is $b$. For the screw, glide edge and climb edge, the nonzero component of the Burgers vector is $b_3$, $b_1$, and $b_2$, respectively. The slip plane, where the material displacement $\mathbf{u}$ experiences a discontinuity, has been chosen as $y=0$ in all cases. In Eq.\ (\ref{eq:betastars}), index $i$ refers to the components of the Burgers vector, and index $j$ to the normal to the plane of discontinuity. Hence, the Burgers vector of the screw is parallel to the dislocation line (in this case, the notion of slip plane proceeds from usual considerations about extended loops\cite{HIRT82}), that of the glide edge is orthogonal to the line and contained in the slip plane, whereas that of the climb edge is orthogonal to the slip plane. Figure \ref{fig:fig1} illustrates these three types of Volterra dislocations in which the reader familiar with fracture theory will recognize dislocation counterparts of the three conventional modes of fracture,\cite{KANN85} the correspondence being (mode III---tearing, mode II---sliding, and mode I---opening) $\leftrightarrow$ (screw, glide edge, and climb edge). The two kinds of edges are of distinct nature: with our sign and orientation conventions, the positive glide edge is obtained by compressing the half-space $y>0$ along $Ox$ in the positive $x$ direction, whereas the positive climb edge results from inserting one extra half-plane of atoms along the negative $x$ semi-axis, which requires to ``open" the pre-existing lattice in the $y$ direction (black region in Fig.\ \ref{fig:fig1}c). As is shown below, these differences express themselves in the analytical expressions of the displacement fields. In usual conditions, the climb edge moves by ``climb", an essentially diffusive mode that involves slow migration of atoms and interstitials. In quasi-static situations, climb dislocations are constituents of sessile prismatic loops or Frank partials.\cite{HIRT82} However, Weertman has emphasized their potential importance in fast dynamics as well.\cite{WEER67} This question is addressed in Sec.\ \ref{sec:ced}.
%----------------------------------------------------------------------------
%----------------------------------------------------------------------------

Equations (\ref{eq:betastars}) are particular cases of the more general expression for a dislocation line \cite{MURA87}
\begin{equation}
\label{eq:surf}
\bm{\beta}^*(\mathbf{x})=\mathbf{b}\otimes\mathbf{n}(\mathbf{x})\,\delta_S(\mathbf{x}),
\end{equation}
where $\delta_S$ is a Dirac distribution localized on the discontinuity surface $S$ of normal $\mathbf{n}(\mathbf{x})$.\cite{NOTE2} The relevance of the location of this discontinuity surface to non-uniform motion is now discussed. Relative to the pristine crystalline state, the structural modification generated by the presence of a  dislocation can be seen as the cumulative effect of elastic atomic displacements, produced by the dislocation core and associated to long-range stresses, and of permanent (irreversible) displacements of atoms accompanying dislocation motion from its nucleation location to its current location. The above eigendistortions are associated to the permanent displacements. By definition, the integral of the relative material (``atomistic") displacements,
\begin{equation}
\label{eq:burgloc}
\int_C \mathrm{d}l\,\frac{\partial u_i}{\partial l}(\mathbf{x})=b^{\text{loc}}_i,
\end{equation}
where $C$ is a closed contour surrounding the dislocation, is non-zero, equal to the \emph{local} Burgers vector, and tends to the true Burgers vector as the loop radius goes to infinity.\cite{HIRT82} This non-zero value materializes a discontinuity of the total material displacement in the medium. From a physical standpoint, once the atomic perturbations generated by the dislocation motion have been damped, crystal integrity is restored behind the dislocation, and the permanent displacements are not observable. For this reason, the above integral is contour-independent for contours large enough compared to the core size, and provides no information on the trajectory followed. Alternatively, the dislocation can be considered as constituting the boundary line of the surface $S$ in Eq.\ (\ref{eq:surf}), so that prescribing this surface removes in practice the indeterminacy of the position of the displacement discontinuity. Therefore, in the static limit and for an infinitely thin dislocation line, the surface can be chosen arbitrarily since the location of the dislocation line alone uniquely determines the elastic strains and stresses.

Resolving this arbitrariness by deciding to localize $\beta^*$ on the geometric surface spanned by the dislocation line during its motion puts in information about the trajectory in the problem.\cite{MURA87} For instance in expressions (\ref{eq:betastars}), the dislocation `comes' from $x=-\infty$. While this information is irrelevant in the static case for infinitely thin dislocations, \emph{this is not true any more in fast dynamics, or for dislocations with extended core such as those considered in the Peierls-Nabarro model}. In the first case, the atoms perturbed by a dislocation passing by oscillate and act as wave sources as long as the perturbation is not fully damped; in the second case, the stacking fault that constitutes the core locally stores potential energy, in particular in case of dissociation and emission of one partial dislocation, and acts as a continuous stress source in the zones where the eigenstrain varies much.

%-------------------------------------------------------------------------------
\subsection{Static Peierls-Nabarro equation}
\label{sec:statpn}
%-------------------------------------------------------------------------------
For clarity and further reference, the method to obtain the static PN equation with the Green's function method is briefly reviewed. The total displacement of a screw dislocation, due to Burgers,\cite{BURG39} is given in all reference textbooks (e.g., Refs.~\onlinecite{HIRT82,MURA87}) as
\begin{equation}
\label{eq:u3scfalse}
u_3(\mathbf{x})=\frac{b\phi}{2\pi}=\frac{b}{2\pi}\mathop{\text{arg}}(x+iy)
=\frac{b}{2\pi}\arctan\frac{y}{x},
\end{equation}
where $\phi$ is the polar angle in the $(x,y)$ plane. However, with the principal determination of the arctangent, the latter expression in Cartesian coordinates is incomplete. Imposing a cut on the negative $x$ semi-axis, the correct result instead reads\cite{PELL09}
\begin{equation}
\label{eq:u3ok}
u_3(\mathbf{x})=\frac{b}{2\pi}\arctan\frac{y}{x}+\frac{b}{2}\mathop{\text{sign}}(y)\theta(-x),
\end{equation}
where the distributional part represents the permanent displacement. The relative slip $\eta$ between both sides of the slip plane is
\begin{equation}
\label{eq:gdif1}
\eta(x)=u_3(x,y=0^+)-u_3(x,y=0^-)=b\theta(-x).
\end{equation}
Even though Eq.\ (\ref{eq:u3ok}) is pretty obvious form the first equality in Eq.\ (\ref{eq:u3scfalse}) and the above elementary remark, retrieving Eq.\ (\ref{eq:u3ok}) using the static Green's function proves a useful exercise prior to considering edges and dynamic calculations. Indeed, the calculation is given by Mura (Ref.~\onlinecite{MURA87}, p.\ 17), with again Eq.\ (\ref{eq:u3scfalse}) as a result. While the cause here may reside in the tables used by this author, the correct calculation is reproduced in Appendix \ref{sec:calcstatvis} for definiteness.

The nonzero components of the total distortion are obtained by differentiation of Eq.\ (\ref{eq:u3ok})~:
\begin{subequations}
 \begin{eqnarray}
 \beta_{zx}&=&u_{z,x}\nonumber\\
 &=&\frac{
b}{2\pi}\left[\pi\delta(x)\mathop{\mathrm{sign}}(y)-\frac{y}{x^2+y^2}\right]
-\frac{b}{2}\mathop{\mathrm{sign}}(y)\delta(x)\nonumber\\
&=&-\frac{ b}{2\pi}\frac{y}{x^2+y^2},\\
 \beta_{zy}&=&u_{z,y}=\frac{
b}{2\pi}\frac{x}{x^2+y^2}+b\delta(y)\theta(-x).
\end{eqnarray}
\end{subequations}
In these expressions, use has been made of the identity
$\arctan x+\arctan 1/x=(\pi/2)\mathop{\text{sign}}x$, from which follows the derivative $(\arctan 1/x)'=\pi\delta(x)-1/(1+x^2)$. The standard textbook expressions of the elastic strain follow from $\varepsilon^e=\mathop{\text{sym}}(\beta_{ij}-\beta^*_{ij})$. It should be noted that distributional parts cancel out, and are absent from the latter expressions, consistently with the fact that a static strain does not depend on history (see previous section). Stresses are obtained by multiplying the (shear) strain components by $2\mu$, as\cite{HIRT82}
\begin{equation}
\label{eq:deplvis}
\sigma_{zx}=-\frac{\mu
b}{2\pi}\frac{y}{x^2+y^2},\qquad \sigma_{zy}=\frac{\mu b}{2\pi}\frac{x}{x^2+y^2}.
\end{equation}
To construct the PN equation, information on the slip plane is reintroduced by computing the stress at $y=0^\pm$. Thus
\begin{subequations}
\label{eq:sressvis}
\begin{eqnarray}
\sigma_{zx}(x,0^\pm)&=&-\frac{\mu b}{2\pi}\lim_{y\to
0}\frac{y}{x^2+y^2}=-\frac{\mu
b}{2}\mathop{\text{sign}}(y)\delta(x),\nonumber\\
&&\\
\label{eq:sigzy}
\sigma_{zy}(x,0)&=&\frac{\mu b}{2\pi}\lim_{y\to
0}\frac{x}{x^2+y^2}=\frac{\mu b}{2\pi}\mathop{\text{p.v.}}\frac{1}{x}.
\end{eqnarray}
\end{subequations}
where $\mathop{\text{p.v.}}$ stands for the principal value. Given expression (\ref{eq:gdif1}) of the differential slip, the stress [Eq.\ (\ref{eq:sigzy})] produced by the dislocation on its slip plane is rewritten as the convolution product
\begin{equation}
\label{eq:sigzy2}
\sigma_{zy}(x,0)=-\frac{\mu}{2\pi}\mathop{\text{p.v.}}
\int\mathrm{d}x'\,\frac{\eta'(x')}{x-x'}.
\end{equation}
This expression now holds for any core shape function $\eta(x)$. Adding an applied resolved shear stress $\sigma^a(x)$ to Eq.\ (\ref{eq:sigzy2}), and balancing their sum by the $b$-periodic pullback force of atomic origin which derives from the stacking fault $\gamma$ potential, hereafter denoted by $f(\eta)$, the static PN equation for the screw is obtained:
\begin{equation}
\label{eq:pnvis}
-\frac{\mu}{2\pi}\mathop{\text{p.v.}}\int_{-\infty}^{+\infty}
\mathrm{d}x'\,\frac{\eta'(x')}{x-x'}+\sigma^a(x)=f'\bigl(\eta(x)\bigr).
\end{equation}
At this point, boundary conditions express the history of dislocation formation; typically $\eta(-\infty)=\eta_0+b$ and
$\eta(+\infty)=\eta_0$ for a single dislocation coming from $x=-\infty$ (Ref.\ \onlinecite{ROSA01})
or $\eta(\pm\infty)=\eta_0$ for a dipole,\cite{NABA47,MOVC98} where $\eta_0$ is the homogeneous solution such that $\sigma_a=f'(\eta_0)$.

In the glide edge case, the nonzero displacement components are $u_x(x,y)$ and $u_y(x,y)$. The Fourier integrals in Mura's method are slightly more complicated, but similar to that for the screw. One finds\cite{EPAPS}
\begin{subequations}
\label{eq:ufieldge-stat}
\begin{eqnarray}
\label{eq:ufieldge-statx}
u_x(x,y)&=&\frac{b}{4\pi}\frac{1}{1-\nu}\frac{xy}{x^2+y^2}\\
&&\qquad{}+\frac{b}{2\pi}
\mathop{\mathrm{arctan}}\frac{y}{x}
+\frac{b}{2}\mathop{\text{sign}}(y)\theta(-x),\nonumber\\
\label{eq:ufieldge-staty}
u_y(x,y)&=&\frac{b}{4\pi}\frac{1}{1-\nu}\frac{y^2}{x^2+y^2}\\
&&\qquad{}-\frac{b}{8\pi}\frac{1-2\nu}{1-\nu}
\log\left[\epsilon_g^2\left(x^2+y^2\right)\right],\nonumber
\end{eqnarray}
\end{subequations}
where $\epsilon_g$ is of order the inverse of half the system size. The sole difference between the present approach and classical results is the presence of the additional distributional term $(b/2)\mathop{\mathrm{sign}}(y)\theta(-x)$ in $u_x$. We recall that because of one divergent integral, $u_y$ is determined only up to an additive constant that blows up as $\epsilon_g\to 0$. For this reason, different equivalent forms of $u_y$ are found in the literature.\cite{BURG39,KOEH41,ESHE49,READ53} This complication, linked to torsion, is well-documented (see Ref.~\onlinecite{HIRT82} p.\ 78). Equation (\ref{eq:ufieldge-staty}) is the form obtained by Eshelby and Mura.\cite{ESHE49,MURA87} Analogous expressions for the climb edge can be written as:
\begin{subequations}
\label{eq:ufieldgec-stat}
\begin{eqnarray}
\label{eq:ufieldgec-statx}
u_x(x,y)&=&-\frac{b}{4\pi}\frac{1}{1-\nu}\frac{x^2}{x^2+y^2}\\
&&\qquad{}+\frac{b}{8\pi}\frac{1-2\nu}{1-\nu}
\log\left[\epsilon_c^2\left(x^2+y^2\right)\right],\nonumber\\
\label{eq:ufieldgec-staty}
u_y(x,y)&=&-\frac{b}{4\pi}\frac{1}{1-\nu}\frac{xy}{x^2+y^2}\\
&&\qquad{}-\frac{b}{2\pi}
\mathop{\mathrm{arctan}}\frac{x}{y}+\frac{b}{4}\mathop{\text{sign}}(y),
\nonumber
\end{eqnarray}
\end{subequations}
where $\varepsilon_c=\varepsilon_g \exp 1/(1-2\nu)$.\cite{EPAPS} When rotated clockwise by a angle $\pi/2$, i.e.\ subjected to substitutions $(x,y)\to (-y,x)$ and $(u_x,u_y)\to (-u_y,u_x)$,  Eqs.\ (\ref{eq:ufieldgec-stat}) become identical to Eqs.\ (\ref{eq:ufieldge-stat}), up to differences in $\epsilon_c$ and in the distributional part. These differences are mathematical renderings of the different nature of glide and climb edges emphasized in Sec.\ \ref{sec:dperm}, though both lead (in the static case only) to identical elastic strains and stresses characteristic of an edge dislocation.

The ensuing static PN equation for the glide edge, for which the resolved stress is $\sigma_{xy}$, is identical to Eq.\ (\ref{eq:pnvis}) save for a prefactor $1/(1-\nu)$ in front of the integral and for the definition of $\eta(x)\equiv u_x(x,0^+)-u_x(x,0^-)$. The static PN equation for the climb edge, driven by a tensile $\sigma_{yy}$ stress component is formally the same as that for the glide edge, but with now $\eta(x)\equiv u_y(x,0^+)-u_y(x,0^-)$. However in the latter case the relevant specialized lattice potential, linked to the introduction of interstitials, has not been properly defined to date to this author's knowledge.

Thus, in all three cases the distributional term plays no part because it does not show up in the elastic strain. The situation markedly changes in dynamics.

%%%%%%%%%%%%%%%%%%%%%%%%%%%%%%%%%%%%%%%%%%%%%%%%%%%%%%%%%%%%%%%%%%%%%%%%%%%%%%%%
\section{\label{sec:dynPN}Dynamic Peierls-Nabarro equation}
%%%%%%%%%%%%%%%%%%%%%%%%%%%%%%%%%%%%%%%%%%%%%%%%%%%%%%%%%%%%%%%%%%%%%%%%%%%%%%%%
The dynamic calculation is quite analogous to the above procedure, using the dynamic Green's function [Eq.\ (\ref{eq:gdyn})] instead of the static one. The main difference is that time-dependent distributional contributions, which generalize the above static ones, are no more irrelevant and provide important contributions to the PN dynamic equation. The difficulty mainly resides in computing cumbersome Fourier integrals. One approach could consist in using the Cagniard-de Hoop method, as proposed by Markenscoff and Clifton\cite{MARK81} to address dynamic dislocation problems. However, using ``brute force" and reference tables of integrals proved more expedient for the case at hand. The edge cases require one key integral that we could not find in tables, which is computed by means of a differential equation. Appendix \ref{sec:calcdyn} contains a detailed sketch of these calculations. Hereafter, $|\mathbf{x}|_2^2=x^2+y^2$.

%-------------------------------------------------------------------------------
\subsection{Principle}
%-------------------------------------------------------------------------------
The dynamic PN equation is obtained in the following manner. First, one computes the elementary stress field produced by a time-dependent eigendistortion $\beta_{ij}^*(\mathbf{x},t)$ representing an ``instantaneous" Volterra dislocation located at $x=y=0$, and present at $t=0$. We take $\beta_{ij}^*(\mathbf{x},t)$ equal to any of Eqs.\ (\ref{eq:betastars}), multiplied by the Dirac impulse $\delta(t)$. Eshelby and more recent works\cite{MARK80} instead consider elementary Volterra dislocations proportional to $\theta(-t)$, but the present approach simplifies the calculation of the stresses in the perspective of obtaining a PN equation. \textit{Mutatis mutandis}, we then follow Eshelby\cite{ESHE49} by appealing to the identity:
\begin{eqnarray}
\label{eq:eshid}
\eta(x,t)&=&\eta(+\infty,t)\\
&-&\int{\mathrm{d}\tau}{\mathrm{d}x'}\theta\bigl(-(x-x')\bigr)\delta(t-
\tau)\frac{\partial \eta}{\partial x}(x',\tau).\nonumber
\end{eqnarray}
Here and below, double integrals over space and time have implicit bounds $\pm\infty$ unless otherwise stated and are denoted by a single integral sign. The integral term expresses the spectrum of instantaneous Volterra dislocations associated to the core shape function $\eta$. No dislocation is associated to the homogeneous slip $\eta(+\infty,t)$. Invoking linear superposition, if $\sigma^{\textrm{elem}}(x,t)$ is the shear stress on the slip plane generated by a Volterra dislocation $b\theta(-x)\delta(t)$, the shear stress generated by the continuous slip $\eta(x,t)=u_i(x,y=0^+)-u_i(x,y=0^-)$ reads, by Eq.\ (\ref{eq:eshid}),
\begin{equation}
\label{eq:sigsigelem}
\sigma(x,t)=-\frac{1}{b}\int{\mathrm{d}\tau}{\mathrm{d}x'}\sigma^{\textrm{elem}}(x-x',t-\tau)
\frac{\partial \eta}{\partial x}(x',\tau).
\end{equation}
An applied inhomogeneous stress in the bulk moreover produces a stress $\sigma^a(x,t)$ on the slip plane, to be added to Eq.\ (\ref{eq:sigsigelem}). Balancing the resulting expression by the pull-back stress yields the dynamic PN equation,
\begin{eqnarray}
&&\hspace{-1.5cm}-\frac{1}{b}\int{\mathrm{d}\tau}{\mathrm{d}x'}
\sigma^{\textrm{elem}}(x-x',t-\tau)
\label{eq:dynpn}
\frac{\partial \eta}{\partial x}(x',\tau)\nonumber\\
&&\hspace{2cm}{}+\sigma_a(x,t)=f'\bigl(\eta(x,t)\bigr),
\end{eqnarray}
where $\eta(\pm\infty,t)$ is such that $\sigma^a(\pm\infty,t)=f'\bigl(\eta(\pm\infty,t)\bigr)$.
We now proceed to determine  $\sigma^{\textrm{elem}}$ for the different kinds of dislocations.

%-------------------------------------------------------------------------------
\subsection{Screw dislocations}
%-------------------------------------------------------------------------------
Then, the displacement associated to the instantaneous screw reads (see Appendix \ref{sec:calcdynv})
\begin{eqnarray}
u_z(\mathbf{x},t)&=&\frac{b
c_{\text{S}}}{2\pi}\Bigl[\frac{x y}{\left(c_{\text{S}}^2
t^2-y^2\right)}\frac{\theta(c_{\text{S}}t-|\mathbf{x}|_2)}{\sqrt{c_{\text{S}}^2
t^2-|\mathbf{x}|_2^2}}\nonumber\\
\label{eq:ufields}
&&{}+\pi\mathop{\mathrm{sign}}(y)\theta(-x)\delta(c_{\text{S}}t-|y|)\Bigr].
\end{eqnarray}
In this expression, the Dirac term is the dynamic counterpart of the static distributional term in Eq.\ (\ref{eq:u3ok}) and represents a wave leaving the slip plane orthogonally to it. The associated elementary shear on the slip plane $y=0$ follows as
\begin{eqnarray}
\sigma^{\text{elem}}(x,t)&\equiv&
\sigma_{zy}(x,y=0,t)\nonumber\\
&=&\lim_{y\to 0}\mu\left[\frac{\partial u_z}{\partial
y}(x,y,t)-\beta^*_{zy}(x,y,t)\right].
\end{eqnarray}
Introducing the kernel
\begin{equation}
\label{eq:kxt} K(x,t)=\frac{x}{2c_{\text{S}}
t^2}\frac{\theta(c_{\text{S}}t-|x|)}{\sqrt{c_{\text{S}}^2
t^2-x^2}},
\end{equation}
one directly finds:
\begin{equation}
\sigma^{\text{elem}}_{zy}(x,y=0,t)=\frac{\mu b}{\pi}K(x,t)-\frac{\mu b}{2c_{\text{S}}}\theta(-x) \delta'(t).
\end{equation}
Applying Eq.\ (\ref{eq:sigsigelem}) produces
\begin{eqnarray}
\sigma_{zy}(x,t)&=&-\frac{\mu}{\pi}\int\mathrm{d}\tau\,\mathrm{d}x'\,K(x-x',t-\tau)\frac{\partial
\eta}{\partial
x}(x',\tau)\nonumber\\
&&\hspace{-12ex}{}+\frac{\mu}{2c_{\text{S}}}\int\mathrm{d}\tau\,\mathrm{d}x'\,\theta\bigl(-(x-x')\bigr)\delta'(t-\tau)\frac{\partial
\eta}{\partial x}(x',\tau).
\end{eqnarray}
Assuming that $\partial \eta/\partial t(+\infty,t)=0$, the second integral reduces to
\begin{eqnarray}
&&\hspace{-2cm}\int\mathrm{d}\tau\,\mathrm{d}x'\,\theta\bigl(-(x-x')\bigr)\delta'(t-\tau)\frac{\partial \eta}{\partial
x}(x',\tau)\nonumber\\
&=&\int_x^{+\infty}\mathrm{d}x'\,\frac{\partial^2 \eta}{\partial
x\partial t}(x',t)=-\frac{\partial \eta}{\partial t}(x,t),
\end{eqnarray}
which gives the time-dependent stress
\begin{eqnarray}
\label{eq:tdstress}
\sigma_{zy}(x,t)&=&-\frac{\mu}{\pi}\int\mathrm{d}\tau\,\mathrm{d}x'\,K(x-x',t-\tau)\frac{\partial
\eta}{\partial x}(x',\tau)\nonumber\\
&&\hspace{2.5cm}{}-\frac{\mu}{2c_{\text{S}}}\frac{\partial
\eta}{\partial t}(x,t).
\end{eqnarray}
Hence, from Eq.\ (\ref{eq:dynpn}), the dynamic PN equation for the screw is
\begin{eqnarray}
\label{eq:dynpns}
&&\hspace{-4em}{}-\frac{\mu}{\pi}\int\mathrm{d}\tau\,\mathrm{d}x'\,K(x-x',t-\tau)\frac{\partial
\eta}{\partial x}(x',\tau)\nonumber\\
&&{}-\frac{\mu}{2c_{\text{S}}}\frac{\partial
\eta}{\partial t}(x,t)+\sigma_a(x,t)=f'\bigl(\eta(x,t)\bigr).
\end{eqnarray}
Apart from the presence of the driving stress $\sigma_a$, this equation differs from Eshelby's [Eq.\ (21) of Ref.~\onlinecite{ESHE53}] by the instantaneous term proportional to $\partial \eta/\partial t$. This term should be replaced here and henceforth by $\mods\partial\mode \widetilde{\eta}/\partial t$, where $\widetilde{\eta}(x,t)=\eta(x,t)-\eta(+\infty,t)$, whenever boundary conditions require $\partial \eta/\partial t(+\infty,t)\not=0$.

%-------------------------------------------------------------------------------
\subsection{Glide edge dislocation}
%-------------------------------------------------------------------------------
Dynamic displacement fields for the instantaneous glide edge are derived in Appendix \ref{sec:calcdyncg}. They read
%%%%% BEGIN WIDETEXT %%%%%%%%%%%%%%%%%%%%%%%%%%%%%%%%%%%%%%%%%%%%%%%%%
\begin{widetext}
\begin{subequations}
\label{eq:ufieldge}
\begin{eqnarray}
u_x(\mathbf{x},t)&=&\frac{b c_{\text{S}}}{2\pi}\theta(t) \left\{
\frac{2xy}{|\mathbf{x}|_2^4}
\left[\frac{c_{\text{S}}}{c_{\text{L}}}\frac{2c_{\text{L}}^2
t^2-|\mathbf{x}|_2^2}{\sqrt{c_{\text{L}}^2
t^2-|\mathbf{x}|_2^2}}\theta(c_{\text{L}}t-|\mathbf{x}|_2)-\frac{2c_{\text{S}}^2
t^2-|\mathbf{x}|_2^2}{\sqrt{c_{\text{S}}^2
t^2-|\mathbf{x}|_2^2}}\theta(c_{\text{S}}t-|\mathbf{x}|_2)\right]
\right.\nonumber\\
\label{eq:ufieldgex}
&&\left.\hspace{5cm}{}
+ \frac{x y}{\left(c_{\text{S}}^2
t^2-y^2\right)}\frac{\theta(c_{\text{S}}t-|\mathbf{x}|_2)}{\sqrt{c_{\text{S}}^2
t^2-|\mathbf{x}|_2^2}}
+\pi\mathop{\mathrm{sign}}(y)\theta(-x)\delta(c_{\text{S}}t-|y|)
\right\}\\
u_y(\mathbf{x},t)&=&\frac{b c_{\text{S}}}{2\pi}\theta(t) \left\{
\frac{2}{|\mathbf{x}|_2^4}\left[\frac{c_{\text{S}}}{c_{\text{L}}}\frac{x^2|\mathbf{x}|_2^2
-c_{\text{L}}^2 t^2(x^2-y^2)}{\sqrt{c_{\text{L}}^2
t^2-|\mathbf{x}|_2^2}}\theta(c_{\text{L}}t-|\mathbf{x}|_2)
%\right.\right.\nonumber\\
\label{eq:ufieldgey}
%&&\left.\left. {}
-\frac{x^2|\mathbf{x}|_2^2 -c_{\text{S}}^2 t^2(x^2-y^2)}{\sqrt{c_{\text{S}}^2
t^2-|\mathbf{x}|_2^2}}\theta(c_{\text{S}}t-|\mathbf{x}|_2)\right]
\right.\nonumber\\
&&{}\hspace{11cm}\left.{}+
\frac{\theta(c_{\text{S}}t-|\mathbf{x}|_2)}{\sqrt{c_{\text{S}}^2t^2-|\mathbf{x}|_2^2}}
\right\}
\end{eqnarray}
\end{subequations}
The corresponding expressions for the distortions and stresses are easy to compute but lengthy so that only $\sigma_{xy}(x,y=0)$, the relevant stress for the PN equation, is reproduced here. Using $\sigma_{xy}=\mu(\beta^e_{xy}+\beta^e_{yx})=\mu(u_{x,y}+u_{y,x}-\beta^*_{xy})$
with $\beta^*_{xy}=b\theta(-x)\delta(y)\delta(t)$ yields
\begin{equation}
\label{eq:siggyzer}
\sigma_{xy}^{\text{elem}}(x,y=0,t)=\frac{\mu b}{\pi}\left[K_1(x,t)+\frac{\partial K_2}{\partial
x}(x,t)\right]-\frac{\mu b}{2 c_{\text{S}}}\theta(-x)\delta'(t),
\end{equation}
where the kernels are:
\begin{subequations}
\label{eq:noyauxcg}
\begin{eqnarray}
\label{eq:noyauxcg1} K_1(x,t)&=&\frac{2c_{\text{S}}}{x^3}
\left[\frac{c_{\text{S}}}{c_{\text{L}}}\frac{2c_{\text{L}}^2
t^2-x^2}{\sqrt{c_{\text{L}}^2
t^2-x^2}}\theta(c_{\text{L}}t-|x|)-\frac{2c_{\text{S}}^2
t^2-x^2}{\sqrt{c_{\text{S}}^2
t^2-x^2}}\theta(c_{\text{S}}t-|x|)\right]
%\nonumber\\
%&&
+\frac{x}{2c_{\text{S}}
t^2}\frac{\theta(c_{\text{S}}t-|x|)}{\sqrt{c_{\text{S}}^2
t^2-x^2}},\\
\label{eq:noyauxcg2}
K_2(x,t)&=&\frac{c_{\text{S}}}{2}\frac{\theta(c_{\text{S}}t-|x|)}{\sqrt{c_{\text{S}}^2t^2-x^2}}.
\end{eqnarray}
\end{subequations}
To arrive at Eq.\ (\ref{eq:siggyzer}), the prescription $\theta(t=0)=1/2$ was used. The highly singular contribution $\partial K_2/\partial x$ in Eq.\ (\ref{eq:siggyzer}) is a distribution that should be used be means of integration by parts. Proceeding as for the screw, the following dynamic PN equation is obtained:
\begin{eqnarray}
\label{eq:pncg}
-\frac{\mu}{\pi}\int\mathrm{d}\tau\,\mathrm{d}x'\,K_1(x-x',t-\tau)\frac{\partial
\eta}{\partial
x}(x',\tau)&-&\frac{\mu}{\pi}\int\mathrm{d}\tau\,\mathrm{d}x'\,K_2(x-x',t-\tau)\frac{\partial^2
\eta}{\partial x^2}(x',\tau)
%\nonumber\\
%&&{}
-\frac{\mu}{2 c_{\text{S}}}\frac{\partial \eta}{\partial
t}(x,t)\nonumber\\
&&\hspace{5cm}{}+\sigma^a(x,t)=f'\bigl(\eta(x,t)\bigr).
\end{eqnarray}
\end{widetext}
%%%%% END WIDETEXT %%%%%%%%%%%%%%%%%%%%%%%%%%%%%%%%%%%%%%%%%%%%%%%%%%%
Remarkably, this equation features a convolution with the second derivative $\partial^2
\eta/\partial x^2$ that was not present for the screw. Whereas the term $\partial \eta/\partial x$, maximal at the dislocation center, is linked to the position of the dislocation, the second-derivative term provides contributions from leading and trailing regions of the core and thus has bearings on the dislocation width.

%-------------------------------------------------------------------------------
\subsection{Climb edge dislocation}
\label{sec:ced}
%-------------------------------------------------------------------------------
%----------------------------------------------------------------------------
Because in principle climb dislocation move diffusively,\cite{HIRT82} the question of their ``relativistic" velocity regime might appear irrelevant. It was however put forward some decades ago in a stationary context by Ang and Williams\cite{ANG59} and Weertman.\cite{WEER67b,WEER69} Weertman advocated that since the atomic spacing across a stacking fault may differ from that across an unfaulted plane, an out-of-plane component of the Burgers vector, of the climb type, ought to be associated to partial dislocations that bound stacking faults on their glide plane.\cite{WEER67b} This component would then move in concert with the in-plane dislocation components, with the fast velocity of the latter and in a diffusionless fashion. Dissociation of a perfect dislocation into partials indeed leaves a possibility that leading and trailing partials have out-of-plane components of opposite sign, leading to an overall asymmetric pattern of the dissociated dislocation. In fact, asymmetric dissociated non-planar cores have recently been observed in molecular-dynamics simulations in copper, in a dynamical context where non-planarity, greatly enhanced by a high applied stress, arises in conjunction with fast motion.\cite{MORD06} A link to climb components might be that non-planar cores can be modeled by using ``core field" corrections,\cite{CLOU09} obtained by prescribing an effective asymmetric dipolar distribution of Burgers vectors that includes out-of-plane components near the core, as a model to nonlinear effects.\cite{GEHL72} In this case however, the climb components are ``ad hoc" and ought not be attributed the physical character of lattice Burgers vectors. If this connection proved true in the dynamical case, such corrections could be derived using spatial derivatives of the displacement field of climb components, although we shall not pursue in this direction. On the other hand, isolated regular climb components may enter the ``dislocation part" of disconnections, a category of defects with a step associated to interfaces between grains or phases, which can move in a diffusionless fashion, e.g., in diffusionless transformation such as twinning or martensitic transformations.\cite{HIRT96} Transonic twinning dislocations were reported in atomistic simulations.\cite{GUMB99b} For all these reasons, we find it worthwhile to write down the dynamic PN equation of a climb component. In the case of interfacial dislocations, and because we deal with isotropic media, our calculation only concerns situations where the phases in presence are of same equivalent isotropic elastic moduli (twinning, notably).

The dynamic displacement field of a climb component is obtained as in the glide case with no additional complications. We only quote the result
%%%%% BEGIN WIDETEXT %%%%%%%%%%%%%%%%%%%%%%%%%%%%%%%%%%%%%%%%%%%%%%%%%
\begin{widetext}
\begin{subequations}
\label{eq:ufieldce}
\begin{eqnarray}
u_x(\mathbf{x},t)&=&\frac{b c_{\text{S}}}{2\pi}\theta(t) \left\{
\frac{2}{|\mathbf{x}|_2^4}\left[\frac{c_{\text{S}}}{c_{\text{L}}}
\frac{x^2|\mathbf{x}|_2^2-c_{\text{L}}^2 t^2(x^2-y^2)}{\sqrt{c_{\text{L}}^2
t^2-|\mathbf{x}|_2^2}}\theta(c_{\text{L}}t-|\mathbf{x}|_2)
%\right.\right.\nonumber\\
%&&\left.\left. {}
-\frac{x^2|\mathbf{x}|_2^2-c_{\text{S}}^2 t^2(x^2-y^2)}{\sqrt{c_{\text{S}}^2
t^2-|\mathbf{x}|_2^2}}\theta(c_{\text{S}}t-|\mathbf{x}|_2)\right]
\right.\nonumber\\
\label{eq:ufieldcex}
&&\hspace{9cm}{}\left.
+\frac{c_{\text{S}}}{c_{\text{L}}}
\left(\frac{c_{\text{L}}^2}{c_{\text{S}}^2}-2\right)
\frac{\theta(c_{\text{L}}t-|\mathbf{x}|_2)}
{\sqrt{c_{\text{L}}^2t^2-|\mathbf{x}|_2^2}}\right\}\\
u_y(\mathbf{x},t)&=&\frac{b c_{\text{S}}}{2\pi}\theta(t) \left\{
-\frac{2 x y}{|\mathbf{x}|_2^4}
\left[\frac{c_{\text{S}}}{c_{\text{L}}}\frac{2c_{\text{L}}^2
t^2-|\mathbf{x}|_2^2}{\sqrt{c_{\text{L}}^2
t^2-|\mathbf{x}|_2^2}}\theta(c_{\text{L}}t-|\mathbf{x}|_2)-\frac{2c_{\text{S}}^2
t^2-|\mathbf{x}|_2^2}{\sqrt{c_{\text{S}}^2
t^2-|\mathbf{x}|_2^2}}\theta(c_{\text{S}}t-|\mathbf{x}|_2)\right]
\right.\nonumber\\
\label{eq:ufieldcey}
&&\hspace{4cm}\left.{}-\frac{c_{\text{L}}}{c_{\text{S}}} \frac{x y}{\left(c_{\text{L}}^2
t^2-x^2\right)}\frac{\theta(c_{\text{L}}t-|\mathbf{x}|_2)}{\sqrt{c_{\text{L}}^2
t^2-|\mathbf{x}|_2^2}}
+\pi\frac{c_{\text{L}}}{c_{\text{S}}}
\mathop{\mathrm{sign}}(y)\theta(-x)\delta(c_{\text{L}}t-|y|)
\right\}.
\end{eqnarray}
\end{subequations}
\end{widetext}
%%%%% END WIDETEXT %%%%%%%%%%%%%%%%%%%%%%%%%%%%%%%%%%%%%%%%%%%%%%%%%%%
Writing $\sigma_{yy}$, the resolved stress of the climb component, as:
\begin{equation}
\sigma_{yy}=\mu\left[\frac{c_{\text{L}}^2}{c_{\text{S}}^2}
\left(u_{y,y}-\beta^*_{yy}\right)+\left(\frac{c_{\text{L}}^2}{c_{\text{S}}^2}
-2\right)u_{x,x}\right],
\end{equation}
its value on the plane $y=0$ reads:
\begin{eqnarray}
\hspace{-2em}\sigma_{yy}(x,y=0,t)&=&\frac{\mu b}{\pi}\left[K_1(x,t)+\frac{\partial K_2}{\partial
x}(x,t)\right]\nonumber\\
\label{eq:sigcxzer}
&&\hspace{1cm}{}-\frac{\mu b}{2}\frac{c_{\text{L}}}{c_{\text{S}}^2}\theta(-x)\delta'(t),
\end{eqnarray}
with the kernels
\begin{subequations}
\label{eq:noyauxcc}
\begin{eqnarray}
\label{eq:noyauxcc1} K_1(x,t)&=&-\frac{2c_{\text{S}}}{x^3}
\left[\frac{c_{\text{S}}}{c_{\text{L}}}\frac{2c_{\text{L}}^2 t^2-x^2}{\sqrt{c_{\text{L}}^2
t^2-x^2}}\theta(c_{\text{L}}t-|x|)\right.\nonumber\\
&&\hspace{-2.5cm}\left.{}-\frac{2c_{\text{S}}^2 t^2-x^2}{\sqrt{c_{\text{S}}^2
t^2-x^2}}\theta(c_{\text{S}}t-|x|)\right]+\frac{c_{\text{L}} x}{2c_{\text{S}}^2
t^2}\frac{\theta(c_{\text{L}}t-|x|)}{\sqrt{c_{\text{L}}^2
t^2-x^2}},\\
\label{eq:noyauxcc2}
K_2(x,t)&=&\frac{c_{\text{S}}^2}{2c_{\text{L}}}
\left(\frac{c_{\text{L}}^2}{c_{\text{S}}^2}-2\right)^2
\frac{\theta(c_{\text{L}}t-|x|)}{\sqrt{c_{\text{L}}^2t^2-x^2}}.
\end{eqnarray}
\end{subequations}
The corresponding dynamic PN equation is of the form (\ref{eq:pncg}), where now $\eta(x)\equiv u_y(x,0^+)-u_y(x,0^-)$,
and where the coefficient of the third (instantaneous) term on the left-hand side (lhs) of Eq.\ (\ref{eq:pncg}), namely, $\mu/(2 c_{\text{S}})$, should be replaced by $\mu c_{\text{L}}/(2 c_{\text{S}}^2)$ according to Eq.\ (\ref{eq:sigcxzer}). This case however remains somewhat formal, for lack of available proper definitions of the associated pull-back force $f'(\eta)$.

%-------------------------------------------------------------------------------
\subsection{Static limit}
\label{sec:static}
%-------------------------------------------------------------------------------
A first independent check of the above results consists in computing from Eqs.\ (\ref{eq:ufields}), (\ref{eq:ufieldgex}), (\ref{eq:ufieldgey}), (\ref{eq:ufieldcex}) and (\ref{eq:ufieldcey}) the following ``static" displacement field:
\begin{equation}
\label{eq:timeint}
\mathbf{u}(\mathbf{x})=\lim_{t\to\infty}\int_{-\infty}^t \mathrm{d}\tau\, \mathbf{u}(\mathbf{x},\tau).
\end{equation}
It is easily found that this integral applied to Eq.\ (\ref{eq:ufields}) gives Eq.\ (\ref{eq:u3ok}) back, and that Eqs.\ (\ref{eq:ufieldge-statx}) and (\ref{eq:ufieldge-staty}) are retrieved with $\epsilon_g=1/(2c_{\text{S}}t)\to 0$ by applying it to Eqs.\ (\ref{eq:ufieldge}). Likewise, the static fields [Eqs.\ (\ref{eq:ufieldgec-statx}) and
(\ref{eq:ufieldgec-staty})] of the climb edge are retrieved from Eq.\ (\ref{eq:ufieldce}), with the following scaling parameter in the logarithm:
$\epsilon_c=\varepsilon_g\left(e^{1/2}c_{\text{L}}/c_{\text{ S}}\right)^{c_{\text{L}}^2/c_{\text{S}}^2-1}$.
For both ``edges", the logarithmic divergence at large sizes is replaced by a divergence at large times, the true static regime being reached when $\epsilon$ becomes of order the inverse system size. Remark in passing that the dynamic ratio $\epsilon_c/\epsilon_g$ found here is different from its static value $\epsilon_c/\epsilon_g=\exp 1/(1-2\nu)=\exp \left(c_{\text{L}}^2/c_{\text{S}}^2-1\right)$ (see Sec.\ \ref{sec:statpn}), owing to differences in the limiting process employed, unless $c_{\text{L}}=e^{1/2}c_{\text{S}}$. The next time-dependent correction in the asymptotic expansion at large times of the integral in Eq.\ (\ref{eq:timeint}) is of order $O(1/t^2)$.

%-------------------------------------------------------------------------------
\subsection{Stationary limit: Weertman's equations}
\label{sec:weert}
%-------------------------------------------------------------------------------
A less trivial independent check consists in computing the stationary limit of the obtained dynamic PN equations. In the stationary regime where the dislocation moves with constant velocity $v$, an ansatz $\eta(x,t)=\eta(x-vt)$ should apply. It is observed that consistency with this ansatz requires the applied stress $\sigma_a(x,t)$ to be either a constant, or a front moving with same velocity of the type $\sigma_a(x,t)=\sigma_a(x-vt)$.
Since a stress front necessarily propagates with one of the sound velocities, the dislocation velocity $v$ is then either equal to this sound velocity---if the glide plane is aligned with the propagation direction of the front or greater---if the glide plane is inclined with respect to this direction. Thus, a stationary propagating front can only involve transonic or supersonic dislocations, and in this case $\sigma_a(x-vt)$ prescribes the velocity.

Under any of these two conditions, it is demonstrated in Appendix \ref{sec:verifw} for the screw and the glide edge (the climb edge is left to the reader) that Weertman's equations\cite{WEER69} are retrieved in the following form, which encompasses all regimes (subsonic, transonic for edge dislocations, and supersonic):
\begin{eqnarray}
&&\hspace{-1cm}{}-\frac{\mu}{\pi}A(v)\mathop{\text{p.v.}}
\int\mathrm{d}x'\,\frac{\eta'(x')}{x-x'}+\mu
B(v)\,\eta'(x)\nonumber\\
\label{eq:pnstat}
&&\hspace{3cm}{}+\sigma_a(x)=f'\bigl(\eta(x)\bigr),
\end{eqnarray}
where for screw dislocations,
\begin{subequations}
\label{eq:abv}
\begin{eqnarray}
\label{eq:abva}
A(v)&=&\frac{1}{2}\sqrt{1-v^2/c_{\text{S}}^2}\,\theta(1-|v|/c_{\text{S}}|),
\\
\label{eq:abvb}
B(v)&=&\mathop{\text{sign}}(v)\frac{1}{2}\sqrt{v^2/c_{\text{S}}^2-1}\,\theta(|v|/c_{\text{S}}-1),
\end{eqnarray}
\end{subequations}
for glide edge dislocations (see also Ref.~\onlinecite{ROSA01}),
\begin{subequations}
\label{eq:abv-ge}
\begin{eqnarray}
A(v)&=&2
\left(\frac{c_{\text{S}}}{v}\right)^2\left[\beta_1\theta(1-|v|/c_{\text{L}})
-\frac{\beta_3^4}{\beta_2}\theta(1-|v|/c_{\text{S}})\right],
\nonumber\\
\label{eq:abva-ge}
\\
B(v)&=&2
\left(\frac{c_{\text{S}}}{v}\right)^2\left[\beta_1\theta(|v|/c_{\text{L}}-1)
+\frac{\beta_3^4}{\beta_2}
\theta(|v|/c_{\text{S}}-1)\right]\nonumber\\
\label{eq:abvb-ge}
&&\hspace{4.5cm}{}\times\mathop{\mathrm{sign}}(v);
\end{eqnarray}
\end{subequations}
and for climb edge dislocations,
\begin{subequations}
\label{eq:abv-ce}
\begin{eqnarray}
A(v)&=&2
\left(\frac{c_{\text{S}}}{v}\right)^2\left[\beta_2\theta(1-|v|/c_{\text{S}})
-\frac{\beta_3^4}{\beta_1}\theta(1-|v|/c_{\text{L}})\right],
\nonumber\\
\label{eq:abva-ce}
\\
\label{eq:abvb-ce}
B(v)&=&2
\left(\frac{c_{\text{S}}}{v}\right)^2\left[\beta_2\theta(|v|/c_{\text{S}}-1)
+\frac{\beta_3^4}{\beta_1}
\theta(|v|/c_{\text{L}}-1)\right]\nonumber\\
&&\hspace{4.5cm}{}\times\mathop{\mathrm{sign}}(v).
\end{eqnarray}
\end{subequations}
These coefficients are expressed in terms of the quantities  $\beta_i=|1-(v/c_i)^2|^{1/2}$, with $c_1=c_{\text{L}}$, $c_2=c_{\text{S}}$, and
$c_3=\sqrt{2}c_{\text{S}}<c_1$.\cite{ROSA01}

%%%%%%%%%%%%%%%%%%%%%%%%%%%%%%%%%%%%%%%%%%%%%%%%%%%%%%%%%%%%%%%%%%%%%
\section{Concluding remarks}
%%%%%%%%%%%%%%%%%%%%%%%%%%%%%%%%%%%%%%%%%%%%%%%%%%%%%%%%%%%%%%%%%%%%%
To summarize, dynamic extensions of the Peierls-Nabarro equation were derived for screw and edge dislocations (of the \emph{glide} and \emph{climb} types) using the Green's function method popularized by Mura,\cite{MURA87} and the Eshelby-type trick of using identity (\ref{eq:eshid}). Besides the instantaneous term that shows up in these equations, the origin of which was traced to a missing distributional term in the displacements, an unexpected feature of the dynamic PN equations is a term involving a convolution with the second space derivative of the displacement jump in both edge cases. The obtained equations formally cover all velocity regimes, as indicated by their stationary limits. Leaving their solution to future work, we conclude with some remarks.

Technically, this result was arrived at by using elementary Volterra solutions proportional to $\delta(t)$, which simplifies calculations. Our expressions for the dynamic stresses induced by the continuous displacements $\eta$ considerably differ from previous results. Consider for instance a moving screw Volterra dislocation at time-varying position $\xi(t)$, starting from rest at $t=0$, represented by the function $\eta(x,t)=b\theta(\xi(t)-x)$ with $\xi(t)=0$ for $t\leq 0$. The time-dependent distortion $u_{z,y}$ generated by such a dislocation has been computed by Markenscoff.\cite{MARK80} Her result [also Eq.\ (1) of Ref.\ \onlinecite{MARK08}] consists of a sum of two integrals, the second one being extremely singular on the glide plane $y=0$, added to a term that compensates for the static field of the dislocation at rest prior to motion. The singularity that develops in her second integral in the limit $y\to 0$ greatly complicates the obtention of the stress on the glide plane. On the contrary, this stress is readily deduced from our Eq.\ (\ref{eq:tdstress}). One obtains for $t>0$:
\begin{eqnarray}
\label{eq:stressy0}
\frac{2\pi\sigma_{xz}}{\mu b}&=&
\frac{1}{c_{\text{S}}}\int_0^t\frac{\text{d}\tau}{(t-\tau)^2}\frac{\overline{v}\,\theta(1-|\overline{v}|)}{\sqrt{1-\overline{v}^2}}-\frac{\pi}{c_{\text{S}}}\delta(\xi-x)\dot{\xi}\nonumber\\
&&{}+\frac{1}{x}\left[1-\sqrt{1-\overline{v}_0\mods^2\mode}\theta(1-|\mods\overline{v}_0\mode|)\right],
\end{eqnarray}
where we wrote for brevity
\begin{equation}
\overline{v}(x,t,\tau)=\frac{x-\xi(\tau)}{c_{\text{s}}(t-\tau)}
\end{equation}
and $\overline{v}_0(x,t)=\overline{v}(x,t,0)$. The last term in Eq.\ (\ref{eq:stressy0}) stems from an explicit integration over times $\tau<0$, and expresses the progressive erosion of the static field within a shell of radius $|x|\leq c_{\text{s}}t$. Note that Eq.\ (\ref{eq:stressy0}) remains extremely singular: besides the Dirac term, the integral over $\tau$ is ill-defined at $\tau=t$. As in the formalism of Markenscoff and co-workers, these singularities arise because the Volterra dislocation is of null width, and can be regularized by using smoother core functions. However, they do not arise in the same fashion. Although the physical and mathematical contents of both approaches ought to be identical, comparing them explicitly proves difficult. For this reason, we found useful to give straightforward independent checks by detailing in the appendix the steps leading to Weertman's equations. It should be noted that the latter derivation does not require separate consideration of the different sonic regimes, contrary to previous works.

Next, the instantaneous term $-[\mu/(2 c_{\text{S}})](\partial\eta/\partial t)(x,t)$, absent from Eshelby's dynamic PN equation for screws, that we obtain in the dynamic equations, is of dissipative nature. It accounts for instantaneous losses by \emph{shear} wave emission transverse to the slip plane as the dislocation advances. In opposition, the nonlocal kernels represent the waves on the slip plane (of the shear type for screws, and of the shear \emph{and} longitudinal types for edges) that determine the core shape. In the \emph{subsonic} steady state, transverse radiative losses in Weertman's equations are exactly compensated by the energy that flows to the core.\cite{ROSA01} This is the meaning of the compensation of terms proportional to $v$ that occurs in the calculations of Appendix \ref{sec:verifw}.

Moreover, the participation of shear waves only to the instantaneous loss term of the screw and glide edges is a consequence of their in-plane character. As the instantaneous term in the dynamic PN equation for the climb makes clear, longitudinal waves too would be emitted by an additional out-of-plane component of the Burgers vector, leading to further dissipation. In this connection, Gumbsch and Gao\cite{GUMB99} already noted that an out-of-plane component would add some drag to the energetically favorable stationary radiation-free transonic regime\cite{ESHE49} for glide edges that occurs at $v=c_3$. This can be seen from Weertman's equations. Indeed, radiative losses are proportional to $B(v)$ in Eqs.\ (\ref{eq:abvb}), (\ref{eq:abvb-ge}) and (\ref{eq:abvb-ce}),\cite{ROSA01} and whereas $B(c_3)=0$ for the glide edge, this term is non-zero for the climb. For lack of proper knowledge about it, we preferred not to conclude on the form of the pull-back force in the case of the PN equation for a climb component, but potentials for glide and climb edge components of the same dislocation ought be coupled in some manner.

Furthermore, although admitting arbitrary velocities, it is not clear to us at present whether the dynamic PN equations should or not require the same additional regularization as Weertman's, their stationary limit, to produce solutions that behave correctly. As they stand Weertman's equations are indeed known to be defective, for two different reasons.\cite{ROSA01} First, in presence of a homogeneous applied stress, these equations admit no single-dislocation solution due to absence of dissipation in the subsonic regime $v<c_{\text{S}}$ where $B(v)=0$. This problem finds its origin in the static PN equation.\cite{MOVC98} To correct it, a phenomenological additional drag term must be prescribed to account for losses of lattice origin. \cite{ROSA01} Evidently, such a term can be added as well in the form $-\alpha (\partial\eta/\partial t)(x,t)$ ($\alpha$ being some drag coefficient) to the lhs of our Eqs.\ (\ref{eq:dynpns}) and (\ref{eq:pncg}), thereby ``renormalizing" the already present loss term, but at the risk of excessively damping sounds waves.\cite{PELL09} The second reason is that for a supersonic transition to take place, ``relativistic" core contraction---a feature of the solutions to Weertman's equations that plausibly carries over to some extent to the instationary regime--- must be forbidden below some microscopic scale to prevent energy from becoming infinite.\cite{ESHE53} In the stationary limit a phenomenological implementation of this constraint consists in curing Weertman's equations by adding a smoothing gradient term,\cite{ROSA01} which provides a connection with the spatial-dispersion effects alluded to in the introduction. Recently, a simple device has been proposed allowing one to overcome both problems at the same time. It essentially consists in a suitable coarse-grained reinterpretation of the PN equation in the static case.\cite{PELL09} A similar procedure could be applied to the dynamic equations. However, it might occur that consideration of full dynamical behavior alleviate the need for such regularizations.

Finally, anisotropy has important consequences on the stress/velocity dependence of dislocation motion, owing to the presence of three sound waves in anisotropic media.\cite{MARI04} The present theory could be extended to this case by appealing to available elementary anisotropic dynamic solutions for displacements.\cite{ANIS}

% To force column breaking here
%\vspace{1cm}
%%%%%%%%%%%%%%%%%%%%%%%%%%%%%%%%%%%%%%%%%%%%%%%%%%%%%%%%%%%%%%%%%%%%%%%%%%%%%%%%%
%%%%%%%%%%%%%%%%%%%%%%%%%%%%%%%%%%%%%%%%%%%%%%%%%%%%%%%%%%%%%%%%%%%%%%%%%%%%%%%%%
\begin{acknowledgments}
The author thanks G.\ Z\'erah \mods{for} \mode having aroused his interest in dislocations, and C.\ Denoual and  R.\ Madec for stimulating discussions.
\end{acknowledgments}

%%%%%%%%%%%%%%%%%%%%%%%%%%%%%%%%%%%%%%%%%%%%%%%%%%%%%%%%%%%%%%%%%%%%%%%
\appendix
%%%%%%%%%%%%%%%%%%%%%%%%%%%%%%%%%%%%%%%%%%%%%%%%%%%%%%%%%%%%%%%%%%%%%%%
%%%%%%%%%%%%%%%%%%%%%%%%%%%%%%%%%%%%%%%%%%%%%%%%%%%%%%%%%%%%%%%%%%%%%%%
\section{STATIC DISPLACEMENTS BY THE GREEN'S FUNCTION METHOD}
%%%%%%%%%%%%%%%%%%%%%%%%%%%%%%%%%%%%%%%%%%%%%%%%%%%%%%%%%%%%%%%%%%%%%%%
\label{sec:calcstatvis}
This section examines only the calculation for the Volterra screw dislocation, as an illustration of how distributional parts emerge from otherwise standard Fourier integrals. For completeness, like calculations for the glide and climb edges are provided in a separate document.\cite{EPAPS}
From (\ref{eq:elast}) and with $\beta^*_{ij}$ given by Eq.\ (\ref{eq:vis}), one has
\begin{equation}
\tau_{ij}=C_{ijkl}\beta^*_{kl}=\mu\beta^*_{32}\left(\delta_{i3}\delta_{j2}+\delta_{i2}\delta_{j3}\right)
\end{equation}
so that with the static Green's function [Eq.\ (\ref{eq:gstat})],
\begin{eqnarray}
[G_{ij}k_k\tau_{kj}](\mathbf{k},t)&=&\frac{1}{
k}\Bigl[\hat{k}_3\delta_{i2}+\hat{k}_2\delta_{i3}\nonumber\\
&&{}-\frac{1}{(1-\nu)}\hat{k}_i
\hat{k}_3 \hat{k}_2\Bigr]\beta^*_{32}(\mathbf{k},t).
\end{eqnarray}
Then, specializing Eq.\ (\ref{eq:udyn}) to the static case by carrying out the time integration [Eq.\ (\ref{eq:timeint})] in the first place,
\begin{equation}
\label{eq:integralscrew}
u_z(x,y)=b\int\frac{\mathrm{d}k_1\,\mathrm{d}k_2}{(2\pi)^2}\frac{e^{i(k_1
x+k_2 y)}}{k_1+i\epsilon}\frac{k_2}{k_1^2+k_2^2}.
\end{equation}
To arrive at this integral, the FT of $\beta^*_{kl}$ was carried out with the help of the (one-dimensional) FT of $\theta(-x)$, which evaluates to $i/(k_1+i\epsilon)$ with $\epsilon\to 0^+$. In Eq.\ (\ref{eq:integralscrew}) The integral over $k_2$ is done first, so as to account for the prescription $\epsilon\to 0$ in the remaining integral over $k_1$. By contour integration,
\begin{equation}
\label{eq:intermediaire1}
\int\frac{\mathrm{d}k_2}{2\pi}\,\frac{k_2\,
e^{i k_2 y}}{k_1^2+k_2^2}=\frac{i}{2}\mathop{\text{sign}}(y)e^{-|k_1||y|},
\end{equation}
and the remaining integral over $k_1$ is `folded' on the positive semi-axis with a change of variables before letting $\epsilon\to 0$. This leads to the integral
\begin{eqnarray}
&&\hspace{-1cm}\int_0^{+\infty}\frac{\mathrm{d}k_1}{2\pi}\,\frac{e^{i k_1
x}}{k_1+i\epsilon}e^{-|k_1||y|}
=i\int_0^{+\infty}\frac{\mathrm{d}k_1}{\pi}e^{- k_1
|y|}\nonumber\\
&&\times\left[\frac{k_1}{k_1^2+\epsilon^2}\sin(k_1
x)-\frac{\epsilon}{k_1^2+\epsilon^2}\cos(k_1 x)\right],\nonumber\\
\label{eq:dirac}
&=&i\int_0^{+\infty}\frac{\mathrm{d}k_1}{\pi}e^{- k_1
|y|}\left[\frac{\sin(k_1 x)}{k_1}-\pi\delta(k_1)\right]
\end{eqnarray}
The way the Dirac distribution arises in Eq.\ (\ref{eq:dirac}) makes clear that the prescription $\int_0^{+\infty}\mathrm{d}k_1\,\delta(k_1)=1/2$ holds. Moreover (G.R.\ 3.941-1),
\begin{eqnarray}
&&\int_0^{+\infty}\frac{\mathrm{d}k_1}{k_1}\,e^{- k_1 |y|}\sin(k_1
x)\nonumber\\
&&\hspace{-0.5cm}{}=\mathop{\text{sign}}(x)\int_0^{+\infty}\frac{\mathrm{d}k_1}{k_1}\,e^{-
k_1\frac{|y|}{|x|}}\sin(k_1)=\mathop{\text{sign}}(x)\tan^{-1}\frac{|x|}{|y|},\nonumber\\
&&
\end{eqnarray}
so that
\begin{eqnarray}
&&\hspace{-1cm}\int_0^{+\infty}\frac{\mathrm{d}k_1}{2\pi}\,\frac{e^{i k_1
x}}{k_1+i\epsilon}e^{-|k_1||y|}\nonumber\\
\label{eq:intermediaire6}
&&{}=-i\left[\theta(-x)+\frac{1}{\pi}\mathop{\text{sign}}(x)\mathop{\mathrm{arctan}}\frac{|y|}{|x|}\right].
\end{eqnarray}
Multiplying by the factor $(i/2)\mathop{\text{sign}}(y)$ coming from
(\ref{eq:intermediaire1}) eventually yields
\begin{eqnarray}
&&\hspace{-1cm}\int\frac{\mathrm{d}k_1\,\mathrm{d}k_2}{(2\pi)^2}\frac{e^{i(k_1
x+k_2 y)}}{k_1+i\epsilon}\frac{k_2}{k_1^2+k_2^2}\nonumber\\
\label{eq:intermediaire3}
&&{}=\frac{1}{2\pi}
\mathop{\mathrm{arctan}}\frac{y}{x}+\frac{1}{2}\mathop{\text{sign}}(y)\theta(-x),
\end{eqnarray}
whence expression (\ref{eq:u3ok}) of $u_z$. The edge cases are addressed by
similar means.\cite{EPAPS}

%%%%%%%%%%%%%%%%%%%%%%%%%%%%%%%%%%%%%%%%%%%%%%%%%%%%%%%%%%%%%%%%%%%%%%%
%%%%%%%%%%%%%%%%%%%%%%%%%%%%%%%%%%%%%%%%%%%%%%%%%%%%%%%%%%%%%%%%%%%%%%%
\section{DYNAMIC DISPLACEMENTS}
\label{sec:calcdyn}
%%%%%%%%%%%%%%%%%%%%%%%%%%%%%%%%%%%%%%%%%%%%%%%%%%%%%%%%%%%%%%%%%%%%%%%
\subsection{Screw dislocation}
\label{sec:calcdynv}
%-------------------------------------------------------------------------------
The instantaneous screw is generated by the eigendistortion of nonzero component
$\beta^*_{zy}(\mathbf{x},t)$ $=$ $b\,\delta(y)\theta(-x)\delta(t)$. With now $k=(k_1^2+k_2^2)^{1/2}$  and using Eq.\ (\ref{eq:udyn}), the displacement takes on the form
\begin{eqnarray}
u_z(\mathbf{x},t)&=&-i\int_{-\infty}^{+\infty}\frac{\mathrm{d}^3k}{(2\pi)^3}
[G_{3j}k_k\tau_{jk}](\mathbf{k},t)e^{i\mathbf{k}\cdot\mathbf{x}}\nonumber\\
&=&b\,
c_{\text{S}}\theta(t)I^{(1)}\mods(x,y,t)\mode,
\end{eqnarray}
where the following integral was introduced:
\begin{eqnarray}
&&\hspace{-1cm}I^{(1)}(x,y,t)=\int\frac{\mathrm{d}k_1\,\mathrm{d}k_2}{(2\pi)^2}\frac{\sin(c
k t)}{k_1+i\epsilon}\hat{k}_2\, e^{i(k_1 x+k_2
y)}\nonumber\\
\label{eq:int2}
&&=-i\frac{\partial}{\partial
y}\int\frac{\mathrm{d}k_1}{2\pi}\frac{e^{i k_1
x}}{k_1+i\epsilon}\int\frac{\mathrm{d}k_2}{2\pi} \frac{\sin\left(c
t k \right)}{k} e^{i k_2 y}.
\end{eqnarray}
In this expression, the inner integral over $k_2$ is (G.R.\ 3.876-1):
\begin{eqnarray}
&&\int\frac{\mathrm{d}k_2}{2\pi} \frac{\sin\left(c
t k\right)}{k} e^{i k_2 y}\nonumber\\
\label{eq:int1}
&&\qquad{}=\frac{1}{2}J_0\left(|k_1|(c^2
t^2-y^2)^{1/2}\right)\theta(ct-|y|).
\end{eqnarray}
For $ct>|y|$, going to the limit $\epsilon\to 0$ as in Eq.\ (\ref{eq:dirac}), the remaining integral is (G.R.\ 6.693-7):
\begin{eqnarray}
&&-i\int\frac{\mathrm{d}k_1}{2\pi}\frac{e^{i k_1
x}}{k_1+i\epsilon}J_0\left(|k_1|(c^2 t^2-y^2)^{1/2}\right)\nonumber\\
&=&\int_0^\infty\frac{\mathrm{d}k_1}{\pi}\left[\frac{\sin(k_1
x)}{k_1}-\pi\delta(k_1)\right]J_0\left(k_1(c^2
t^2-y^2)^{1/2}\right)\nonumber\\
&=&-\frac{1}{2}+\frac{1}{\pi}\int_0^\infty\frac{\mathrm{d}u}{u}\sin\left(u
x(c^2 t^2-y^2)^{-1/2}\right)J_0(u)
\nonumber\\
%\end{eqnarray}
%\begin{eqnarray}
&=&-\frac{1}{2}+\frac{1}{2}\mathop{\text{sign}}(x)\theta(|\mathbf{x}|_2^2-c^2t^2)
\nonumber\\
&&{}+\frac{1}{\pi}\arcsin\left(\frac{x}{\sqrt{c^2
t^2-y^2}}\right)\theta(c^2t^2-|\mathbf{x}|_2^2)\nonumber\\
&=&\left[\frac{1}{\pi}\arcsin\left(\frac{x}{\sqrt{c^2
t^2-y^2}}\right)\!-\!\frac{1}{2}\mathop{\text{sign}}(x)\right]\theta(c^2t^2-|\mathbf{x}|_2^2)\nonumber\\
&&\hspace{5.5cm}{}-\theta(-x).
\end{eqnarray}
Multiplying by $(1/2)\theta(ct-|y|)$
according to Eq.\ (\ref{eq:int1}), and differentiating the product with respect to $y$
according to Eq.\ (\ref{eq:int2}) yields
\begin{eqnarray}
I^{(1)}(x,y,t)&=&\frac{1}{2\pi}\Biggl[\frac{x
y}{\left(c^2
t^2-y^2\right)}\frac{\theta(ct-|\mathbf{x}|_2)}{\sqrt{c^2
t^2-|\mathbf{x}|_2^2}}\nonumber\\
\label{eq:int3} &&{}+\pi\mathop{\mathrm{sign}}(y)\theta(-x)\delta(ct-|y|)\Biggr].
\end{eqnarray}
Equation (\ref{eq:ufields}) follows.

%-------------------------------------------------------------------------------
\subsection{Glide edge dislocation}
\label{sec:calcdyncg}
%-------------------------------------------------------------------------------
The only non-zero component of $\beta$ is now
$\beta^*_{12}(\mathbf{x},t)=b\,\delta(y)\theta(-x)\delta(t)$, and
% Then,
%\begin{equation}
$k_k\tau_{kj}=\mu\beta^*_{12}\left(k_1\delta_{j2}+k_2\delta_{j1}\right)$.
%\end{equation}
Thus
\begin{eqnarray}
G_{ij}k_k\tau_{kj}\!&=&\!\theta(t)c_{\text{S}}^2\left[\frac{1}{c_{\text{S}}}\sin(c_{\text{S}} k
t)\!\left(\hat{k}_1\delta_{i2}\!+\!\hat{k}_2\delta_{i1}\!-\!2\hat{k}_i \hat{k}_1\hat{k}_2\right)\right.\nonumber\\
&&\hspace{0.8cm}\left.{}+\frac{2}{c_{\text{L}}}\sin(c_{\text{L}} k t)\hat{k}_i
\hat{k}_1 \hat{k}_2\right]\beta^*_{xy}.
\end{eqnarray}
Setting $k=(k_1^2+k_2^2)^{1/2}$, the non-zero components of $\mathbf{u}$ are obtained as:
\begin{subequations}
\begin{eqnarray}
&&u_x(\mathbf{x},t)=-i\int_{-\infty}^{+\infty}\frac{\mathrm{d}^3k}{(2\pi)^3}
[G_{1j}k_k\tau_{jk}](\mathbf{k},t)e^{i\mathbf{k}\cdot\mathbf{x}}\nonumber\\
&&=b c_{\text{S}}^2\theta(t)
\int\frac{\mathrm{d}k_1\,\mathrm{d}k_2}{(2\pi)^2} e^{i(k_1 x+k_2
y)}\left\{\frac{\sin(c_{\text{S}} k
t)k_2}{c_{\text{S}}k(k_1+i\epsilon)}\right.\nonumber\\
\label{eq:uxapp}
&&\left.\quad{}+2\left[\frac{\sin(c_{\text{L}}
k t)}{c_{\text{L}}}-\frac{\sin(c_{\text{S}} k
t)}{c_{\text{S}}}\right]\frac{k_1 k_2}{k^3}\right\}
\end{eqnarray}
and
\begin{eqnarray}
&&u_y(\mathbf{x},t)=-i\int_{-\infty}^{+\infty}\frac{\mathrm{d}^3k}{(2\pi)^3}
[G_{2j}k_k\tau_{jk}](\mathbf{k},t)e^{i\mathbf{k}\cdot\mathbf{x}}\nonumber\\
&&=b c_{\text{S}}^2\theta(t)
\int\frac{\mathrm{d}k_1\,\mathrm{d}k_2}{(2\pi)^2} e^{i(k_1 x+k_2
y)}\left\{\frac{\sin(c_{\text{S}} k
t)}{c_{\text{S}}k}\right.\nonumber\\
\label{eq:uyapp}
&&\quad\left.{}+2\left[\frac{\sin(c_{\text{L}} k
t)}{c_{\text{L}}}-\frac{\sin(c_{\text{S}} k
t)}{c_{\text{S}}}\right]\frac{k_2^2}{k^3}\right\}.
\end{eqnarray}
\end{subequations}
In these expressions, the limit
$\epsilon\to 0$ was taken wherever possible (cancellation of $k_1$ between numerator and denominator of fractions). Four different types of integrals are involved. The fist one,
$I^{(1)}$, was defined in Eq.\ (\ref{eq:int2}) and computed in Eq.\ (\ref{eq:int3}). The three others ones are (G.R.\ 8.411-5 and 6.671-7)
\begin{subequations}
\label{eq:threeintegrals}
\begin{eqnarray}
\label{eq:i2}
I^{(2)}(x,y,t)\!&=&\!\int\frac{\mathrm{d}k_1\,\mathrm{d}k_2}{(2\pi)^2}\frac{\sin(c
k t)}{k} e^{i(k_1 x+k_2
y)}\\
&&\hspace{-2cm}{}=\int_0^\infty\frac{\mathrm{d}k}{2\pi}\,\sin(c k
t)J_0(k|\mathbf{x}|_2)=\frac{1}{2\pi}
\frac{\theta(ct-|\mathbf{x}|_2)}{\sqrt{c^2t^2-|\mathbf{x}|_2^2}},
\nonumber\\
I^{(3)}(x,y,t)\!&=&\!\int\frac{\mathrm{d}k_1\,\mathrm{d}k_2}{(2\pi)^2}\frac{\sin(c
k t)k_1k_2}{k^3} e^{i(k_1 x+k_2 y)}=\frac{\partial J}{\partial
x},\nonumber\\
\label{eq:i3}
\\
I^{(4)}(x,y,t)\!&=&\!\int\frac{\mathrm{d}k_1\,\mathrm{d}k_2}{(2\pi)^2}\frac{\sin(c
k t)k_2^2}{k^3} e^{i(k_1 x+k_2 y)}=\frac{\partial J}{\partial
y},\nonumber\\
\label{eq:i4}
\end{eqnarray}
\end{subequations}
where the following integral was introduced:
\begin{eqnarray}
J(x,y,t)&=&-i\int\frac{\mathrm{d}k_1\,\mathrm{d}k_2}{(2\pi)^2}\frac{\sin(c
k t)k_2}{k^3} e^{i(k_1 x+k_2
y)}\nonumber\\
&=&\mathop{\text{sign}}(y)
\int_0^\infty\frac{\mathrm{d}k_1}{\pi}\frac{\cos(k_1
x)}{k_1}\nonumber\\
\label{eq:bigjtwo}
&&\hspace{-2cm}{}\times\int_0^\infty\frac{\mathrm{d}q}{\pi}\frac{q\sin(q
k_1|y|)}{(1+q^2)^{3/2}}\sin\left(c t k_1(1+q^2)^{1/2}\right).
\end{eqnarray}
Its expression in polar coordinates shows that $J$ is finite. The last equality in Eq.\ (\ref{eq:bigjtwo}) follows from elementary symmetry considerations and from a change of variables $k_2\to q=k_2/|k_1|$. Consider first the inner integral, and introduce for convenience ($a$ and $b$ are arbitrary positive constants),
\begin{equation}
j(a,b)=\int_0^\infty\frac{\mathrm{d}q}{\pi}\frac{q\sin(b
q)}{(1+q^2)^{3/2}}\sin\left(a(1+q^2)^{1/2}\right).
\end{equation}
This integral is not tabulated for all positive $(a,b)$ pairs (see G.R.\ 3.875-3 for $a<b$). However, one integration by parts over $q$ and the use of
(G.R.\ 3.876-1) show that ($J_0$ is the Bessel function)
\begin{eqnarray}
j(a,b)&=&a \frac{\partial j}{\partial
a}(a,b)\nonumber\\
&&{}+\int_0^\infty\frac{\mathrm{d}q}{\pi}\frac{\cos(b
q)}{(1+q^2)^{1/2}}\sin\left(a(1+q^2)^{1/2}\right)\nonumber\\
&&\hspace{-1cm}{}=a \frac{\partial j}{\partial a}(a,b)+\frac{b}{2}J_0\left((a^2-b^2)^{1/2}\right)\theta(a-b).
\end{eqnarray}
Thus, $j$ is a continuous solution of a homogeneous (resp.\ non-homogeneous) differential equation for $a<b$ (resp.\ $a>b$). This differential equation is solved by variation of constants with condition $j(\infty,b)=0$ (see G.R.\ 6.554-4 for the integration constant). A change of variables then gives
$$
j(a,b)=\frac{a}{2}\left[e^{-b}-b\,\theta(a-b)
\int_0^{(a^2-b^2)^{1/2}}\hspace{-1em}\frac{u
J_0(u)\,\mathrm{d}u}{(u^2+b^2)^{3/2}}\right].
$$
It follows that
\begin{eqnarray}
J(x,y,t)&=&\mathop{\text{sign}}(y)\int_0^\infty\frac{\mathrm{d}k_1}{\pi}\frac{\cos(k_1
x)}{k_1}j(ct
k_1,|y|k_1)\nonumber\\
&=&\frac{ct}{2\pi}\mathop{\text{sign}}(y)\left[\int_0^\infty\mathrm{d}k_1\,\cos(k_1
x)e^{-k_1|y|}\right.\nonumber\\
&&\hspace{-1cm}{}-|y|\,\theta(ct-|y|)
\int_0^{(c^2t^2-y^2)^{1/2}}\hspace{-1em}\frac{u\,\mathrm{d}u}{(u^2+y^2)^{3/2}}\nonumber\\
&&\hspace{-1cm}\left.{}\times\int_0^\infty\mathrm{d}k_1\,\cos(k_1 x)J_0(u k_1)\right],
\end{eqnarray}
that is, with (G.R.\ 3.893-2) and (G.R.\ 6.671-8),
\begin{eqnarray}
J(x,y,t)&=& \frac{c t
y}{2\pi}\left[\frac{1}{x^2+y^2}\right.\nonumber\\
&&\hspace{-2cm}\left.{}-\theta(ct-|\mathbf{x}|_2)
\int_{|x|}^{(c^2t^2-y^2)^{1/2}}\hspace{-1em}\frac{u\,\mathrm{d}u}{(u^2+y^2)^{3/2}}\frac{1}{(u^2-x^2)^{1/2}}
\right],\nonumber\\
&&\hspace{-1.5cm}{}=\frac{c
t}{2\pi}\frac{y}{|\mathbf{x}|_2^2}\left[1-\frac{1}{ct}\sqrt{c^2t^2-|\mathbf{x}|_2^2}\,\theta(ct-|\mathbf{x}|_2)\right].
\end{eqnarray}
Integrals $I^{(3)}$ and $I^{(4)}$ follow from differentiation according to Eqs.\ (\ref{eq:i3}), (\ref{eq:i4}) as
\begin{eqnarray}
I^{(3)}(x,y,t)\!\!&=&\!\!\frac{ct
}{2\pi}\frac{xy}{|\mathbf{x}|_2^4}\left[\frac{2c^2
t^2-|\mathbf{x}|_2^2}{c t \sqrt{c^2
t^2-|\mathbf{x}|_2^2}}\theta(ct-|\mathbf{x}|_2)-2\right],\nonumber\\
I^{(4)}(x,y,t)\!\!&=&\!\!\frac{ct
}{2\pi}\frac{1}{|\mathbf{x}|_2^4}\Biggl[\frac{x^2|\mathbf{x}|_2^2
-c^2 t^2(x^2-y^2)}{c t \sqrt{c^2
t^2-|\mathbf{x}|_2^2}}\theta(ct-|\mathbf{x}|_2)\nonumber\\
&&\hspace{1.3cm}{}+(x^2-y^2)\Biggr].
\end{eqnarray}
Gathering all contributions within Eqs.\ (\ref{eq:uxapp}), (\ref{eq:uyapp}) then yields displacements (\ref{eq:ufieldgex}) and (\ref{eq:ufieldgey}).

%%%%%%%%%%%%%%%%%%%%%%%%%%%%%%%%%%%%%%%%%%%%%%%%%%%%%%%%%%%%%%%%%%%%%%%
\section{STATIONARY LIMIT}
\label{sec:verifw}
%%%%%%%%%%%%%%%%%%%%%%%%%%%%%%%%%%%%%%%%%%%%%%%%%%%%%%%%%%%%%%%%%%%%%%%
\subsection{Screw dislocation}
\label{sec:verifwv}
%-------------------------------------------------------------------------------
Using $\sigma_a(x,t)=\sigma_a(x-vt)$ and the ansatz $\eta(x,t)=\eta(x-vt)$  (see Sec.\ \ref{sec:weert})
in Eq.\ (\ref{eq:dynpn}), one sees that $\eta(x)$ obeys the PN-like equation
\begin{equation}
\label{eq:statpn1}
-\frac{\mu}{\pi}\int\mathrm{d}x'\,K_v(x-x')\eta'(x')+\frac{\mu
v}{2c_{\text{S}}}\eta'(x)+\sigma_a(x)=f'\bigl(\eta(x)\bigr),
\end{equation}
where
\begin{equation}
\label{eq:kvx}
K_v(x)\equiv\int_0^\infty \mathrm{d}t\,
K(x+vt,t)=\int\frac{\mathrm{d}k}{2\pi}e^{ikx} K_v(k)
\end{equation}
which features the space Fourier transform of $K_v(x)$ in the form of a one-sided integral over time,
\begin{equation}
\label{eq:kvk}
K_v(k)\equiv\int_0^\infty \mathrm{d}t\, e^{i k v t}
K(k,t).
\end{equation}
In this expression $K(k,t)$ is the space FT of $K(x,t)$ [given in (\ref{eq:kxt})], which reads
(G.R.\ 3.752-2; $J_1$ is the Bessel function):
\begin{eqnarray}
K(k,t)&=&-\frac{ik}{c_{\text{S}}t^2}\,\theta(t)\,\int_0^{c_{\text{S}}t}{\rm d}x\,  \sqrt{c_{\text{S}}^2
t^2-x^2}\cos(k x)\nonumber\\
&=&-\frac{i\pi}{2t}\theta(t)\, J_1(c_{\text{S}}k t).
\end{eqnarray}
The expression of $K_v(k)$ is evaluated from (\ref{eq:kvk}) with the help of the integrals (G.R.\ 6.693-1 and 6.693-2)
\begin{subequations}
\label{eq:j1ints}
\begin{eqnarray}
&&\hspace{-1cm}\int_0^\infty\frac{\mathrm{d}t}{t}\cos(kvt)J_1(kc_{\text{S}}t)\nonumber\\
\label{eq:j1ints1}
&&\hspace{-0.5cm}{}=\mathop{\text{sign}}(k)\sqrt{1-v^2/c_{\text{S}}^2}\,\theta(1-|v|/c_{\text{S}}),\\
&&\hspace{-1cm}\int_0^\infty\frac{\mathrm{d}t}{t}\sin(kvt)J_1(kc_{\text{S}}t)\nonumber\\
\label{eq:j1ints2}
&&\hspace{-0.5cm}{}=(v/c_{\text{S}})-\mathop{\text{sign}}(v)\sqrt{v^2/c_{\text{S}}^2-1}\,\theta(|v|/c_{\text{S}}-1),
\end{eqnarray}
\end{subequations}
from which:
\begin{eqnarray}
K_v(k)&=&-i\pi\mathop{\text{sign}}(k)
\frac{1}{2}\sqrt{1-v^2/c_{\text{S}}^2}\theta(1-|v|/c_{\text{S}})\nonumber\\
\label{eq:skkvk}
&&\hspace{-1.7cm}{}+\frac{\pi}{2}\left[(v/c_{\text{S}})-
\mathop{\text{sign}}(v)\sqrt{v^2/c_{\text{S}}^2-1}\theta(|v|/c_{\text{S}}-1)\right].
\end{eqnarray}
Since $-i\pi\mathop{\text{sign}}(k)$ is the FT of
$\mathop{\text{p.v.}}1/x$, the Fourier inversion of $K_v(k)$ is immediate as
\begin{eqnarray}
K_v(x)&=&\theta(1-|v/c_{\text{S}}|)\frac{1}{2}\sqrt{1-v^2/c_{\text{S}}^2}\mathop{\text{p.v.}}
\frac{1}{x}\nonumber\\
\label{eq:skkvx}
&&\hspace{-2.3cm}{}+\frac{\pi}{2}\left[(v/c_{\text{S}})-
\mathop{\text{sign}}(v)\sqrt{v^2/c_{\text{S}}^2-1}\,\theta(|v|/c_{\text{S}}-1)\right]\delta(x).
\end{eqnarray}
Putting this expression into (\ref{eq:statpn1}) one sees that the instantaneous terms (proportional to $v$) cancel out mutually.  Weertman's equation (\ref{eq:pnstat}) with coefficients (\ref{eq:abva}) and (\ref{eq:abvb}) follows.
%\vspace{1cm}

%-------------------------------------------------------------------------------
\subsection{Glide edge dislocation}
\label{sec:verifwcg}
%-------------------------------------------------------------------------------
Again using $\sigma_a(x,t)=\sigma_a(x-vt)$ and the ansatz $\eta(x,t)=\eta(x-vt)$
in the dynamic PN equation (\ref{eq:pncg}) for the glide edge, the resulting
stationary equation takes on the form (\ref{eq:statpn1}) where now
\begin{equation}
\label{eq:kvdef}
K_v(x)=\int_0^\infty\mathrm{d}t\,\left[K_1(x+vt,t)-\frac{\partial
K_2}{\partial x}(x+vt,t)\right],
\end{equation}
in which the kernels $K_1$ and $K_2$ are given by (\ref{eq:noyauxcg}). Proceeding as for the screw in Sec.\ \ref{sec:verifwv}, one evaluates first the Fourier transforms of $K_1$ and $\partial K_2/\partial x$ wrt.\ $x$. By means of changes of variable
$x\to u=x/(c_{\text{L}}t)$ and $x\to u=x/(c_{\text{S}}t)$,
and using the fact that $K_1(x,t)$ is odd in $x$ and that $K_2(x,t)$ is even, one gets with (G.R.\ 3.753-5) and (G.R.\ 3.753-2)
\begin{subequations}
\begin{eqnarray}
\label{eq:k1inter1}
K_1(k,t)&=&-\frac{i\pi}{2t}J_1(k c_{\text{S}}t)-\frac{4ic_{\text{S}}^2}{t}\mathop{\mathrm{sign}}(k)\\
&&\hspace{-2cm}{}\times\int_0^1\!\mathrm{d}u
\left[\frac{\sin(|k|c_{\text{L}}t u)}{c_{\text{L}}^2}-\frac{\sin(|k|c_{\text{S}}t u)}{c_{\text{S}}^2}\right]
\frac{2-u^2}{u^3\sqrt{1-u^2}},\nonumber\\
\label{eq:k2kt}
\left[\frac{\partial K_2}{\partial
x}\right](k,t)&=&-i\frac{\pi}{2}kc_{\text{S}} J_0(kc_{\text{S}}t).
\end{eqnarray}
\end{subequations}
The next step consists in obtaining $\int_0^\infty\mathrm{d}t\,e^{ikvt}K_1(k,t)$ in which we write
\begin{equation}
\label{eq:expdev}
e^{ikvt}=\cos(k v t)+i\mathop{\mathrm{sign}}(k)\mathop{\mathrm{sign}}(v)\sin(|k||v|t)
\end{equation}
%\begin{eqnarray}
%\label{eq:k1inter2}
%\int_0^\infty\mathrm{d}t\,e^{ikvt}K_1(k,t)&=&
%\int_0^\infty\mathrm{d}t\,\cos(k v t)K_1(k,t)\nonumber\\
%&&\hspace{-4cm}{}+i\mathop{\mathrm{sign}}(k)\mathop{\mathrm{sign}}(v)
%\int_0^\infty\mathrm{d}t\,\sin(|k||v|t)K_1(k,t).
%\end{eqnarray}
With (\ref{eq:k1inter1}) and (\ref{eq:expdev}), the latter integral involves the following integrals over time, where $c$
stands either for $c_{\text{L}}$ or for $c_{\text{S}}$ (G.R.\ 3.741-1,2):
\begin{subequations}
\label{eq:k1inter3}
\begin{eqnarray}
\label{eq:k1inter31}
\int_0^\infty\frac{\mathrm{d}t}{t}\,\cos(|k|vt)\sin(|k| c t
u)&=&\frac{\pi}{2}\theta(u-|v|/c),\\
\int_0^\infty\frac{\mathrm{d}t}{t}\,\sin(|k||v|t)\sin(|k| c t
u)&=&\frac{1}{4}\log\left(\frac{u+|v|/c}{u-|v|/c}\right)^2.\nonumber\\
\label{eq:k1inter32}
\end{eqnarray}
\end{subequations}
The remaining integrals over $u$ combined with (\ref{eq:k1inter3}) are evaluated using the pair of integrals,
\begin{subequations}
%%%%% BEGIN WIDETEXT %%%%%%%%%%%%%%%%%%%%%%%%%%%%%%%%%%%%%%%%%%%%%%%%%
\begin{widetext}
\begin{eqnarray}
\int_0^1\mathrm{d}\!u\,\frac{\pi}{2}\theta(u-|v|/c)\frac{2-u^2}{u^3\sqrt{1-u^2}}&=&\frac{\pi}{2}\frac{c^2}{v^2}\sqrt{1-v^2/c^2}\,\theta(1-|v|/c),\\
\int_0^1\mathrm{d}\!u\,\frac{1}{4}\log\left(\frac{u+|v|/c}{u-|v|/c}\right)^2
\frac{2-u^2}{u^3\sqrt{1-u^2}}&=&\frac{2c}{|v|}\int_\epsilon^1\frac{\mathrm{d}u}{u^2}+\int_0^1\mathrm{d}\!u\,\left[\log\left(\frac{u+|v|/c}{u-|v|/c}\right)^2
\frac{2-u^2}{4 u^3\sqrt{1-u^2}}-\frac{2c}{|v|}\frac{1}{u^2}\right]\nonumber\\
&&\hspace{-4cm}{}=\frac{2c}{|v|\epsilon}-\frac{2c}{|v|}+\left.\left[\frac{c}{|v| u}-
\frac{\sqrt{1-u^2}}{4u^2}\log\left(\frac{u+|v|/c}{u-|v|/c}\right)^2\right]\right|^1_0-\frac{c}{|v|}\mathop{\text{p.v.}}
\int_0^1\frac{\mathrm{d}\!u}{u^2}\left[1+\frac{v^2}{c^2}\frac{\sqrt{1-u^2}}{u^2-(v/c)^2}\right]\nonumber\\
\label{eq:lastint}
&&\hspace{-4cm}{}=\lim_{x\to
0+}\frac{c^2t}{|v|}\frac{2}{x}-\frac{\pi}{2}\frac{c^2}{v^2}
\sqrt{v^2/c^2-1}\,\theta(|v|/c-1).
\end{eqnarray}
\end{widetext}
%%%%% END WIDETEXT %%%%%%%%%%%%%%%%%%%%%%%%%%%%%%%%%%%%%%%%%%%%%%%%%%%
\end{subequations}
In (\ref{eq:lastint}) the following transformations were applied. The integral is divergent at $u=0$. Its divergent part is extracted first, and expressed in terms of $x$, recalling that $u$ was introduced via the change of variables $u=x/(ct)$. In this form, it is proportional to $c^2$ and cancels out when assembling contributions involving $c_{\text{L}}$ and $c_{\text{S}}$ in the final step of the calculation. Meanwhile, the remaining finite part is integrated by parts, and the spurious singularity at $u=|v|/c$ introduced by this transformation for $|v|<c$ is removed by the principal value prescription. Appealing next to (\ref{eq:j1ints1}), (\ref{eq:j1ints2}) to deal with the Bessel function in (\ref{eq:k1inter1}), these contributions to (\ref{eq:k1inter1}) lead to:
%%%%% BEGIN WIDETEXT %%%%%%%%%%%%%%%%%%%%%%%%%%%%%%%%%%%%%%%%%%%%%%%%%
\begin{widetext}
\begin{eqnarray}
&&\hspace{-1.4cm}\int_0^\infty\mathrm{d}t\,e^{ikvt}K_1(k,t)=-2i\pi\mathop{\mathrm{sign}}(k)
\frac{c_{\text{S}}^2}{v^2}\left[\sqrt{1-v^2/c_{\text{L}}^2}
\,\theta(1-|v|/c_{\text{L}})+\left(\frac{v^2}{4c_{\text{S}}^2}-1\right)
\sqrt{1-v^2/c_{\text{S}}^2}\,\theta(1-|v|/c_{\text{S}})\right]\nonumber\\
&&\hspace{2cm}-2\pi\mathop{\mathrm{sign}}(v)
\frac{c_{\text{S}}^2}{v^2}\left[\sqrt{v^2/c_{\text{L}}^2-1}
\,\theta(|v|/c_{\text{L}}-1)+\left(\frac{v^2}{4c_{\text{S}}^2}-1\right)
\sqrt{v^2/c_{\text{S}}^2-1}\,\theta(|v|/c_{\text{S}}-1)\right]
+\frac{\pi v}{2c_{\text{S}}}.
\end{eqnarray}
Turning now to $K_2$ one finds, using (\ref{eq:k2kt}), (\ref{eq:expdev}) and  (G.R.\ 6.671-7,8)
\begin{equation}
\int_0^\infty\mathrm{d}t\,e^{ikvt}\left[\frac{\partial K_2}{\partial
x}\right](k,t)=-i\frac{\pi}{2}\mathop{\mathrm{sign}}(k)\frac{\theta(1-|v|/c_{\text{S}})}{\sqrt{1-v^2/c_{\text{S}}^2}}+\mathop{\mathrm{sign}}(v)\frac{\pi}{2}
\frac{\theta(|v|/c_{\text{S}}-1)}{\sqrt{v^2/c_{\text{S}}^2-1}}.
\end{equation}
\end{widetext}
%%%%% END WIDETEXT %%%%%%%%%%%%%%%%%%%%%%%%%%%%%%%%%%%%%%%%%%%%%%%%%%%
The contributions of $K_1$, $\partial K_2/\partial x$ are brought back into $K_v(k)$, whose Fourier inversion is immediate as in the screw case, see Eqs.\ (\ref{eq:skkvk}) and (\ref{eq:skkvx}). The result reads
\begin{equation}
%K_v(x)=A(v)\mathop{\text{p.v.}} (1/x)-\pi B(v)\delta(x)+(\pi c_{\text{L}} v)/(2c_{\text{S}}^2)\delta(x),
K_v(x)=A(v)\mathop{\text{p.v.}}\frac{1}{x}-\pi B(v)\delta(x)+\frac{\pi}{2} \frac{v}{\mods c_{\text{S}}\mode}\delta(x),
\end{equation}
with $A(v)$ and $B(v)$ given by (\ref{eq:abva-ge}) and (\ref{eq:abvb-ge}). This brings the present edge version of (\ref{eq:statpn1}) down to Weertman's equation.


\begin{thebibliography}{99}
\bibitem{NOTE0}
That is, up to non-linear effects. If one uses linear elasticity, as in the present paper, $\beta^*$ is allowed to contain nonlinear elastic contributions (see, e.g., Ref.\ \onlinecite{PELL09} for further insight).
\bibitem{NOTE2}
To ease the connection with (\ref{eq:betastars}), our orientation convention for $\mathbf{n}$ in Eq.\ (\ref{eq:surf}) differs from Mura's (Ref.\ \onlinecite{MURA87}).
\end{thebibliography}
\end{document}